# Solving weighted and counting variants of connectivity problems parameterized by treewidth deterministically in single exponential time

Hans L. Bodlaender
Utrecht University
The Netherlands
H.L.Bodlaender@uu.nl

Marek Cygan
University of Warsaw
Poland
cygan@mimuw.edu.pl

Stefan Kratsch*
Max-Planck-Institute
Saarbrücken, Germany
skratsch@mpi-inf.mpg.de

Jesper Nederlof[†]
Utrecht University
The Netherlands
J.Nederlof@uu.nl

November 8, 2012

## Abstract

It is well known that many local graph problems, like Vertex Cover and Dominating Set, can be solved in $2^{\mathcal{O}(\texttt{tw})}|V|^{\mathcal{O}(1)}$ time for graphs $G = (V, E)$ with a given tree decomposition of width $\texttt{tw}$. However, for nonlocal problems, like the fundamental class of connectivity problems, for a long time we did not know how to do this faster than $\texttt{tw}^{\mathcal{O}(\texttt{tw})}|V|^{\mathcal{O}(1)}$. Recently, Cygan et al. (FOCS 2011) presented Monte Carlo algorithms for a wide range of connectivity problems running in time $c^{\texttt{tw}}|V|^{\mathcal{O}(1)}$ for a small constant $c$, e.g., for Hamiltonian Cycle and Steiner tree. Naturally, this raises the question whether randomization is necessary to achieve this runtime; furthermore, it is desirable to also solve counting and weighted versions (the latter without incurring a pseudo-polynomial cost in terms of the weights).

We present two new approaches rooted in linear algebra, based on matrix rank and determinants, which provide deterministic $c^{\texttt{tw}}|V|^{\mathcal{O}(1)}$ time algorithms, also for weighted and counting versions. For example, in this time we can solve TRAVELING SALESMAN or count the number of Hamiltonian cycles. The rank-based ideas provide a rather general approach for speeding up even straightforward dynamic programming formulations by identifying "small" sets of representative partial solutions; we focus on the case of expressing connectivity via sets of partitions, but the essential ideas should have further applications. The determinant-based approach uses the matrix tree theorem for deriving closed formulas for counting versions of connectivity problems; we show how to evaluate those formulas via dynamic programming.

## 1 Introduction

The notion of *treewidth* proved to be an excellent tool for dealing with many NP-hard problems on graphs. In the 1970s and 1980s, several groups of researchers discovered the concept independently. In their fundamental work on graph minors, Robertson and Seymour [45] introduced the notions *treewidth* and *tree decomposition*, and these became the dominant terminology. The notion has been extensively studied and used in various areas of (theoretical and applied) computer science (see for example [7] for a survey). Informally, the notion measures how well a graph can be decomposed in a tree-like manner, resulting in a so-called *tree decomposition*. This is useful since many efficient algorithms solving NP-hard problems on trees generalize to efficient algorithms for graphs with good tree decompositions. Almost without exception these algorithms crucially rely on the *dynamic programming* paradigm, so studying the complexity of problems on graphs of small treewidth amounts to studying the dynamic programming paradigm.

---

*Part of the work was done at Utrecht University supported by the Nederlandse Organisatie voor Wetenschappelijk Onderzoek (NWO), project: 'KERNELS'.

[†]Supported by the Nederlandse Organisatie voor Wetenschappelijk Onderzoek (NWO), project: 'Space and Time Efficient Structural Improvements of Dynamic Programming Algorithms'.

Two influential results are Bodlaender's [8] and Courcelle's theorems [16], showing that if we assume that input graphs have bounded treewidth, then there are linear time algorithms to find an optimal tree decomposition and for each property that can be expressed in monadic second order logic, given such a tree decomposition, to decide if the property holds for the input graph. Algorithmic engineering evaluations of Bodlaender's and Courcelle's algorithms are not encouraging however, due to the large constant factors in the running times, even for small values of the treewidth, e.g., the running time of Bodlaender's algorithm is $O(\mathtt{tw}^{O(\mathtt{tw}^3)} n)$.

Hence a natural question is how much we can optimize the dependence on the treewidth, that is, we aim for a running time of the type $f(\mathtt{tw})n^c$ on graphs with $n$ vertices and a tree decomposition of width $\mathtt{tw}$ of where we first aim at obtaining a small growing (but, since we assume $\mathcal{P} \neq \mathcal{NP}$, exponential) function $f(\mathtt{tw})$ and second a small exponent $c$. For problems with locally checkable certificates, that is, certificates assigning a constant number of bits per node that can be checked by a cardinality check and iteratively looking at all neighborhoods of the input graph[1], it quickly became clear that $f(\mathtt{tw})$ only needs to be single-exponential. See [33, 42] for sample applications to the Independent Set/Vertex Cover problems. From the work of [31] and known Karp-reductions between problems it follows that this dependence cannot be improved to subexponential algorithms unless the Exponential Time Hypothesis (ETH) fails, i.e., unless CNF-SAT has a subexponential algorithm. In [37] it was shown that under a stronger assumption (the so-called Strong Exponential Time Hypothesis, SETH), the current algorithms are optimal even with respect to polynomial jumps, that is, problems with current best running time $f(\mathtt{tw})n^{\mathcal{O}(1)}$ cannot be solved in $f(\mathtt{pw})^{1-\epsilon}n^{\mathcal{O}(1)}$ for positive $\epsilon$ where $f(\mathtt{pw})$ is $2^{\mathtt{pw}}$, $3^{\mathtt{pw}}$ for respectively Independent Set and Dominating Set. While there is no consensus opinion on whether ETH and SETH are true, improving beyond lower bounds proven under either of them is at least as hard as improving the state of the art of CNF-SAT in the respective way.

A natural class of problems that does not have locally checkable certificates are connectivity problems such as Hamiltonian cycle and Steiner Tree (see for example [27, Section 5]), begging the question whether these can be solved within single-exponential dependence on $\mathtt{tw}$ as well. Early work shows that if we assume, in addition to small treewidth, that the input graph is planar, or, more general, avoids a minor, then many connectivity problems have algorithms with single-exponential dependence of $\mathtt{tw}$, exploiting Catalan structures [20]. A positive answer to the question, using a randomized approach was found by Cygan et al. [18] using an approach termed "Cut & Count": It provided a transformation of the natural certificates to "cut-certificates" transforming the connectivity requirement into a locally checkable requirement. The transformation is only valid modulo 2, but by a standard technique [41] introducing randomization, the decision variant can be reduced to the counting modulo 2 variant. This result was considered surprising since in the folklore $2^{\mathcal{O}(\mathtt{tw}\log\mathtt{tw})}n^{\mathcal{O}(1)}$ dynamic programming routines for connectivity problems all information stored seemed needed: Given two partial solutions inducing different connectivity properties, one may very well be extendable to a solution while the other one is not (this resembles the notion of Myhill-Nerode equivalence classes [29]).

---

[1] For example, for the odd cycle transversal problem that asks to make the input graph by removing at most $k$ vertices, a locally checkable certificate would be a solution set combined with a proper two-coloring of the remaining bipartite graph.

The Cut & Count approach is one of the algorithms from the family of dynamic programming based algorithms using a modulo 2 based transformation [4, 6, 36, 35, 38, 46]. These algorithms give the smallest running times currently known, but have several disadvantages compared to traditional dynamic programming algorithms: (a) They are probabilistic. (b) The dependence on the inputs weights in weighted extensions is pseudo-polynomial. (c) They do not extend to counting the number of witnesses. (d) They do not give intuition for the optimal substructure / equivalence classes. An additional disadvantage of the Cut & Count approach of [18], compared to traditional dynamic programming algorithms on tree decompositions, is that their dependence in terms of the input graph is superlinear. Our work shows that each of these disadvantages can be overcome, with two different approaches, both giving deterministic algorithms for connectivity problems that are singly-exponential in the treewidth.

## 1.1 Our contribution

Our contribution consists of two new approaches that resolve the aforementioned issues of the Cut & Count approach, the "Rank based" and "Squared determinant" approaches. They can be used to quickly and deterministically solve respectively weighted and counting versions of problems solved by the Cut & Count approach. Additional advantages of the rank based approach are that it gives a more intuitive insight in the optimal substructure / equivalence classes of a problem and that is has only a linear dependence on the input graph in the running time. The only disadvantage of both approaches when compared to the Cut & Count approach is that the dependence on the treewidth or pathwidth in the running time is slightly worse. However, although we did not manage to overcome it, this disadvantage might not be inherently due to the new methods. Due to the generality of our key ideas, we feel our methods may inspire future work not involving tree decompositions as well.

Although both approaches apply to all problems studied in [18]. We only study (variants of) STEINER TREE, HAMILTONIAN CYCLE and FEEDBACK VERTEX SET in this work in order to prevent losing focus. Our results on these problems are described in Table 1.

|   | Problem | Pathwidth | Treewidth |
|---|---|---|---|
| 1 | WEIGHTED STEINER TREE | $n(1+2^\omega)^{\mathtt{pw}}\mathtt{pw}^{\mathcal{O}(1)}$ | $n(1+2^{\omega+1})^{\mathtt{tw}}\mathtt{tw}^{\mathcal{O}(1)}$ |
| 2 | TRAVELING SALESMAN | $n(2+2^{\omega/2})^{\mathtt{pw}}\mathtt{pw}^{\mathcal{O}(1)}$ | $n(5+2^{(\omega+2)/2})^{\mathtt{tw}}\mathtt{tw}^{\mathcal{O}(1)}$ |
| 3 | $k$-PATH | $n(2+2^{\omega/2})^{\mathtt{pw}}(k+\mathtt{pw})^{\mathcal{O}(1)}$ | $n(5+2^{(\omega+2)/2})^{\mathtt{tw}}(k+\mathtt{tw})^{\mathcal{O}(1)}$ |
| 4 | FEEDBACK VERTEX SET | $n(1+2^\omega)^{\mathtt{pw}}\mathtt{pw}^{\mathcal{O}(1)}$ | $n(1+2^{\omega+1})^{\mathtt{tw}}\mathtt{tw}^{\mathcal{O}(1)}$ |
| 5 | # HAMILTONIAN CYCLE | $\tilde{\mathcal{O}}(6^{\mathtt{pw}}\mathtt{pw}^{\mathcal{O}(1)}n^2)$ | $\tilde{\mathcal{O}}(15^{\mathtt{tw}}\mathtt{tw}^{\mathcal{O}(1)}n^2)$ |
| 6 | # STEINER TREE | $\tilde{\mathcal{O}}(5^{\mathtt{pw}}\mathtt{pw}^{\mathcal{O}(1)}n^3)$ | $\tilde{\mathcal{O}}(10^{\mathtt{tw}}\mathtt{tw}^{\mathcal{O}(1)}n^3)$ |
| 7 | FEEDBACK VERTEX SET | $\tilde{\mathcal{O}}(5^{\mathtt{pw}}\mathtt{pw}^{\mathcal{O}(1)}n^3)$ | $\tilde{\mathcal{O}}(10^{\mathtt{tw}}\mathtt{tw}^{\mathcal{O}(1)}n^3)$ |

Table 1: Table with our results. Rows $1-4$ are discussed in Section 3 while rows $5-7$ are discussed in Section 4. Line 7 is the result of applying the determinant approach to FEEDBACK VERTEX SET (note that we do not solve the counting variant). The symbol $\omega$ denotes the matrix multiplication exponent (currently it is known that $\omega < 2.3727$ due to [47]).

Since one of the main strengths of the treewidth concept seems to be its ubiquity, it is perhaps not surprising that our results improve, simplify, generalize or unify a number of seemingly unrelated results. In all cases, problem instances can be reduced to instances with small tree- or pathwidth using standard techniques. This is further discussed in Subsection 1.2.

We would like to mention that very recently, a subset of the current authors found an efficient rank bound of a partial solution versus partial solution matrix for the Hamiltonian cycle problem. We will use this in the current to get a low running time for the TRAVELING SALESMAN problem.

### 1.1.1 Rank based approach

Our first tool is the rank based method (presented in Section 3). This is based on a very generic idea to improve a dynamic programming algorithm, that we will now outline.

Recall that a dynamic programming algorithm fixes a way to decompose certificates into 'partial certificates', and builds partial certificates in a bottom-up manner while storing only their essential information. Given some language $L \subseteq \{0,1\}^*$, this is typically implemented by a defining an equivalence $\sim$ on partial certificates $x, y \in \{0,1\}^k$ such that $x \sim y$ if $xz \in L \leftrightarrow yz \in L$, for every extension $z \in \{0,1\}^l$. For connectivity problems on treewidth, the number of non-equivalent certificates can be seen to be larger than our running times. For example [44] for a lower bound in communication complexity.

We will use however, that sometimes we can represent the joint essential information for sets of partial certificates more efficiently than naively representing essential information for every partial certificate separately. The rank based approach achieves this as follows: Given a dynamic programming algorithm, consider the matrix $A$ whose rows and columns are indexed by partial certificates, with $A[x, y] = 1$ if and only if $xy \in L$. Then observe that if a set of rows $X \subseteq \{0,1\}^n$ is linearly dependent (modulo 2), any partial certificate $x \in X$ is redundant in the sense that if $xz \in L$ there will be $y \in X$ with $yz \in L$. Hence, the essential information can be reduced to $\operatorname{rk}(A)$ partial certificates.

For getting this approach to work we require an upper bound on the rank of the matrix $\mathcal{M}$ defined as follows: Fix a ground set $U$ and let $p$ and $q$ be partitions of $U$ ($p$ and $q$ represent connectivity induced by partial solutions), define $\mathcal{M}[p, q]$ to be 1 if and only if the meet of $p$ and $q$ is the trivial partition, that is, if the union of the partial solutions induce a connected solution. Although this matrix has dimensions of order $2^{\theta(|U| \lg |U|)}$, we given a simple factorization in $GF(2)$ of matrices with inner dimension $2^{|U|}$ using an idea of [18]. Interestingly, our factorization follows from more generic results in communication complexity [39, 40]. Furthermore, our factorization (Lemma 3.13) can be derived from the much more general setting of communication complexity [39, 40] (or more specifically, Lemma 5.7 from [39]).

### 1.1.2 Squared determinant approach

Our second tool is the squared determinant approach (presented in Section 4). This gives a generic transformation of counting connected objects to a more local transformation. In [18] the existence of such a transformation in GF(2) was already given. For extending this to counting problems, we will need three key insights. The first insight is that (a variant of) Kirchhoff's matrix tree theorem gives a reduction from counting connected objects to computing a sum of determinants. However, we cannot fully control the contribution of a connected object (it will appear to be either 1 or -1). To overcome this we ensure that every connected object contributed exactly once, we compute a sum of *squares of determinants*. The last obstacle is that the computation of a determinant is not entirely local (in the sense that we can verify a contribution by iteratively considering its intersection with all bags) since we have to account for the number of inversions of a permutation in every summand of the determinant. To overcome this, we use that this in fact becomes a local computation once we have fixed the order of the vertices in a clever way.

## 1.2 Relations to previous work

As mentioned before, our results improve, simplify, generalize or unify a number of seemingly unrelated results. We will support this claim in this subsection.

### 1.2.1 Algorithms for problems with small solutions.

Although we do not improve the running time of known algorithms, a significant advantage of our approach is that it unifies several known non-trivial algorithms. We proceed by giving three important examples:

**Finding long paths** In [2], the authors gave a deterministic $2^{\mathcal{O}(k)}|E|\lg|V|$ time algorithm that finds a simple path on $k$ vertices in a graph $G = (V, E)$, resolving the open question whether simple paths of length $\log n$ can be found in polynomial time from [43]. In a follow-up work, [14] this was improved to a $4^{k+\mathcal{O}(\log^3 k)}|V||E|$ time algorithm. All these algorithms use the (arguably) involved construction of perfect hash families, and no $2^{\mathcal{O}(k)}|V|^{\mathcal{O}(1)}$ time deterministic algorithm avoiding the use of such families was previously known.

As a corollary of our technique we can obtain a simple $\mathcal{O}(2^{\mathcal{O}(k)}k^{\mathcal{O}(1)}|V|+|E|)$ time algorithm, or an $\mathcal{O}(4.28^k k^{\mathcal{O}(1)}|V| + |E|)$ with a standard technique from [23] (see also [10]) as follows. Create a depth first search tree of the input graph; if the height of the tree is more than $k$, the longest path from a leaf to the root gives a $k$-path and we are done. Otherwise, there is a path decomposition of width at most $k$, with for every leaf-root path a bag with the vertices on this path. Since this can be performed in linear time (see [10] for details) we can respectively apply a simple variant of the algorithm from Section 3.4 or the algorithm from Theorem 3.20 to obtain the algorithms.

**Finding feedback vertex sets** Many papers (see [18, 13] and the references therein) have been devoted to the FEEDBACK VERTEX SET problem: Given an input graph $G = (V, E)$ and a (small) parameter $k$, can we remove at most $k$ vertices from $G$ in order to obtain a forest? It should be noted that if this is indeed the case, then a tree decomposition of width $k + 1$ of $G$ is easily obtained from a solution $X$ by adding $X$ to all bags of a tree decomposition of the forest $V \setminus X$. Based on this simple observation, deterministic $2^{\mathcal{O}(k)}k^{\mathcal{O}(1)}|V| + |E|$ algorithms are easily obtained by using a constant-factor approximation or the method of iterative compression. We would like to mention that this is very different from previously known deterministic $2^{\mathcal{O}(k)}k^{\mathcal{O}(1)}|V|^{\mathcal{O}(1)}$ algorithms that all rely on a measure-based branching strategy.

**H-minor free graphs** In [19], it was shown that any $H$-minor-free graph either has treewidth bounded by $\mathcal{O}(\sqrt{k})$, or a $2\sqrt{k} \times 2\sqrt{k}$ grid as a minor. Many problems can easily be solved once we know that the input graph has such a grid (for instance a $2\sqrt{k} \times 2\sqrt{k}$ grid as a minor guarantees that the graph does not have a vertex cover smaller than $k$ or does have a $4k$-path).

Thus, we are left with solving the problem with the assumption of bounded treewidth. In the case of problems with a local verification algorithm such as vertex cover, a standard dynamic programming algorithm then gives us a $2^{\mathcal{O}(\sqrt{k})}n^{\mathcal{O}(1)}$ algorithm to check whether a graph has a vertex cover no larger than $k$. In the case of $k$-path, however, the standard dynamic programming routine gives a $2^{\mathcal{O}(\sqrt{k}\log k)}n^{\mathcal{O}(1)}$ complexity. Motivated by this a number of papers showed that the fact that $G$ is $H$-minor free can be exploited to give $2^{\mathcal{O}(\mathtt{tw})}n^{\mathcal{O}(1)}$ time algorithms (see [20] and the references therein). In [18] this result was extended to general graphs at the cost of randomization and a pseudo-polynomial dependence on the input weights, if present. In the current work we resolve the latter issues thereby fully unifying these previous results.

### 1.2.2 Graph classes with small treewidth

Sometimes the input is restricted to a graph class that is guaranteed to have small treewidth by known results. We survey here some connections of our work to previous results that are more tailor made for such a graph class.

**Bounded degree** A number of problems has been studied in the setting were the input graph has bounded degree. For example, the TRAVELING SALESMAN has received considerable attention [5, 22, 26, 32]. Our contribution is a derandomization at the cost of a slightly larger base of the exponential dependence.

Throughout the following, we let $n$ denote the number of vertices of the given graph. Then the following theorem by Fomin et al. reduces the bounded degree setting to small treewidth:

**Theorem 1.1** ([24]). *For any $\epsilon > 0$ there exists an integer $n_\epsilon$ such that for any graph $G$ with $n > n_\epsilon$ vertices,*
$$\mathtt{pw}(G) \leq \frac{1}{6}n_3 + \frac{1}{3}n_4 + \frac{13}{30}n_5 + n_{\geq 6} + \epsilon n,$$
*where $n_i$ is the number of vertices of degree $i$ in $G$ for any $i \in \{3, \ldots, 5\}$ and $n_{\geq 6}$ is the number of vertices of degree at least 6.*

This theorem is constructive, and the corresponding path decomposition (and, consequently, tree decomposition) can be found in polynomial time. Using this result one can get fast determinstic algorithms for all the connectivity problems listed in Subsection 1.3.3 in [18]. In this paper we restrict ourselves to the most studied setting, TRAVELING SALESMAN problem for cubic graphs. In [18] a $1.201^n n^{\mathcal{O}(1)}$ Monte-Carlo algorithm was obtained for Hamiltonian cycle; Very recently in [17], this is improved to a $1.1583^n n^{\mathcal{O}(1)}$ time Monte-Carlo algorithm for Hamiltonian cycle. By combining the ideas of this paper with a result from [17], we obtain a $1.2186^n n^{\mathcal{O}(1)}$ time deterministic algorithm for the TRAVELING SALESMAN problem in Corollary 3.19.

**Planar graphs** Recall from the previous subsection that $n$ denotes the number of vertices of the given graph. Here we begin with a consequence of [25]:

**Proposition 1.2** ([25]). *For any planar graph $G$, $\mathtt{tw}(G) + 1 \leq \frac{3}{2}\sqrt{4.5n} \leq 3.183\sqrt{n}$. Moreover a tree decomposition of such width can be found in polynomial time.*

Using this we immediately obtain the following result for TRAVELING SALESMAN on planar graphs.

**Corollary 1.3.** *There exists an algorithm that solves TRAVELING SALESMAN on planar graphs in $4.28^{3.183\sqrt{n}} n^{\mathcal{O}(1)} = 2^{6.677\sqrt{n}} n^{\mathcal{O}(1)}$ time.*

The best known algorithm so far was the $\mathcal{O}(2^{6.903\sqrt{n}})$-time algorithm of Bodlaender et al. [21]. A $4^{3.183\sqrt{n}} n^{\mathcal{O}(1)}$ randomized algorithm for Hamiltonian cycle was given in [18].

Similarly, we obtain an $\mathcal{O}(2^{6.677\sqrt{n}})$-time deterministic algorithm for $k$-cycle on planar graphs (compare to the $\mathcal{O}(2^{7.223\sqrt{n}})$ time of [21]), and well-behaved $c^{\sqrt{n}}$-time algorithms for the other connectivity problems studied in [18].

## 2 Preliminaries

We use standard graph theory notation. Additionally for a subset of edges $X \subseteq E$ of an undirected graph $G = (V, E)$ by $G[X]$ we denote the subgraph induced by edges and endpoints of $X$, i.e. $G[X] = (V(X), X)$. For $X, Y \subseteq V$ we let $E(X, Y)$ be the set of all edges with one endpoint in $X$ and one in $Y$. For a vertex $v$ and $X \subseteq V$ we denote $\deg_X(v)$ for the number of neighbors of $v$ contained in $X$.

**Partitions and the partition lattice.** Given a base set $U$, we use $\Pi(U)$ for the set of all partitions of $U$. It is known that, together with the coarsening relation $\sqsubseteq$, $\Pi(U)$ gives a lattice, with the minimum element being $\{U\}$ and the maximum element being the partition into singletons. We denote $\sqcap$ for the meet operation and $\sqcup$ for the join operation in this lattice; these operators are associative and commutative. We use $\Pi_2(U) \subset \Pi(U)$ to denote the set of all partitions of $U$ in blocks of size 2, or equivalently, the set of perfect matchings over $U$. Given $p \in \Pi(U)$ we let $\#\texttt{blocks}(p)$ denote the number of blocks of $p$. If $X \subseteq U$ we let $p_{\downarrow X} \in \Pi(X)$ be the partition obtained by removing all elements not in $X$ from it, and analogously we let for $U \subseteq X$ denote $p_{\uparrow X} \in \Pi(X)$ for the partition obtained by adding singletons for every element in $X \setminus U$ to $p$. Also, for $X \subseteq U$, we let $U[X]$ be the partition of $U$ where one block is $\{X\}$ and all other blocks are singletons. If $a, b \in U$ we shorthand $U[ab] = U[\{a, b\}]$. The empty set, vector and partition are all denoted by $\emptyset$.

**Tree decompositions and treewidth.** A *tree decomposition* [45] of a graph $G$ is a tree $\mathbb{T}$ in which each node $x$ has an assigned set of vertices $B_x \subseteq V$ (called a *bag*) such that $\bigcup_{x \in \mathbb{T}} B_x = V$ with the following properties:

- for any $uv \in E$, there exists an $x \in \mathbb{T}$ such that $u, v \in B_x$.

- if $v \in B_x$ and $v \in B_y$, then $v \in B_z$ for all $z$ on the (unique) path from $x$ to $y$ in $\mathbb{T}$.

The *treewidth* $tw(\mathbb{T})$ of a tree decomposition $\mathbb{T}$ is the size of the largest bag of $\mathbb{T}$ minus one, and the treewidth of a graph $G$ is the minimum treewidth over all possible tree decompositions of $G$ (thus the graphs of treewidth one are exactly the forests and trees and that are not also independent sets). Finding a tree decomposition of minimum treewidth is $\mathcal{NP}$-hard [3], but for each fixed integer $k$ there is a linear time algorithm for finding a tree decomposition of width at most $k$ (if it exists) [8]. In this paper, we will always assume that tree decompositions of the appropriate width are given.

Dynamic programming algorithms on tree decompositions are often presented on nice tree decompositions, introduced by Kloks [34]; see also [9, 28] regarding treewidth and dynamic programming. We present a slightly augmented variant that specifies bags/nodes for introducing edges; this was already used by Cygan et al. [18].

**Definition 2.1** (Nice Tree Decomposition). A *nice tree decomposition* is a tree decomposition with one special bag $z$ called the *root* and in which each bag is one of the following types:

- **Leaf bag**: a leaf $x$ of $\mathbb{T}$ with $B_x = \emptyset$.

- **Introduce vertex bag**: an internal vertex $x$ of $\mathbb{T}$ with one child vertex $y$ for which $B_x = B_y \cup \{v\}$ for some $v \notin B_y$. This bag is said to *introduce* $v$.

- **Introduce edge bag**: an internal vertex $x$ of $\mathbb{T}$ labeled with an edge $uv \in E$ with one child bag $y$ for which $u, v \in B_x = B_y$. This bag is said to *introduce* $uv$.

- **Forget bag**: an internal vertex $x$ of $\mathbb{T}$ with one child bag $y$ for which $B_x = B_y \setminus \{v\}$ for some $v \in B_y$. This bag is said to *forget* $v$.

- **Join bag**: an internal vertex $x$ with two child vertices $l$ and $r$ with $B_x = B_r = B_l$.

We additionally require that every edge in $E$ is introduced exactly once.

**Proposition 2.2.** *Given a tree decomposition of some graph $G = (V, E)$, a nice tree decomposition rooted at an empty bag, an introduce edge bag for any choice of $\{u, v\} \in E$ or an forget vertex bag can be computed in $\mathtt{ntw}^{\mathcal{O}(1)}$ time.*

By fixing the root of $\mathbb{T}$, we associate with each bag $x$ in a tree decomposition $\mathbb{T}$ a vertex set $V_x \subseteq V$ where a vertex $v$ belongs to $V_x$ if and only if there is a bag $y$ which is a descendant of $x$ in $\mathbb{T}$ with $v \in B_y$ (recall that $x$ is its own descendant). We also associate with each bag $x$ of $\mathbb{T}$ a subgraph of $G$ as follows:

$$G_x = \Big(V_x, E_x = \{e | e \text{ is introduced in a descendant of } x \}\Big)$$

**Path decompositions and pathwidth.** A *path decomposition* is a tree decomposition where the tree of nodes is restricted to be a path. The *pathwidth* of a graph is the minimum width of all path decompositions. Path decompositions can, similarly as above, be transformed into nice path decompositions, these obviously contain no join bags.

**Further notation.** Given a graph $G = (V, E)$ and a For two integers $a, b$ we use $a \equiv b$ to indicate that $a$ is even if and only if $b$ is even. We use $\mathbb{N}$ to denote the set of all non-negative integers. We use Iverson's bracket notation: if $p$ is a predicate we let $[p]$ be 1 if $p$ if true and 0 otherwise. If $\omega : U \to \{1, \dots, N\}$, we shorthand $\omega(S) = \sum_{e \in S} \omega(e)$ for $S \subseteq U$. For a function $s$ by $s[v \to \alpha]$ we denote the function $s \setminus \{(v, s(v))\} \cup \{(v, \alpha)\}$. Note that this definition works regardless of whether $s(v)$ is already defined or not. We use either $s|_X$ or $s_{|X}$ to denote the vector obtained by restricting the domain to $X$.

## 3 The rank based approach

This section is devoted to a complete presentation of our rank based approach for connectivity problems parameterized by treewidth. At the heart of the approach is a technique for reducing large sets of partial solutions, expressed as weighted partitions (to be defined later), down to smaller sets that maintain the same optimal solution value along with at least one optimal solution (though we point out that this reduction by design cannot maintain all optimal solutions). The main result behind the approach is presented in Section 3.2; it ensures that we can always find a small set of representative partial solutions. To avoid creating a series of ad hoc results for single problems, we introduce a collection of operations on families of weighted partitions, such that our results apply to any dynamic programming (DP) formulation that can be expressed using these operators only (see Section 3.1).

In Sections 3.3, 3.4, and 3.5 we give DP formulations using our operators for STEINER TREE, TRAVELING SALESMAN, and FEEDBACK VERTEX SET respectively. Except for FEEDBACK VERTEX SET, where this turns out to be more involved, the programs are close to naive formulations that give runtimes $2^{\Omega(\mathtt{tw} \cdot \log \mathtt{tw})}$ (note that our operators and language of weighted partitions are nonstandard). At the first read of this section, it is perhaps useful to read only one of the DP formulations, but the different problems solved by them should indicate the versatility of our operators.

Section 3.6 presents the proofs for the main building blocks of the approach (those given in Section 3.2). Finally, Section 3.7 discusses limitations and possible improvements to the results presented in this section.

## 3.1 Operators on sets of weighted partitions

We will now introduce formally what we mean by sets of weighted partitions and followed by a definition of the mentioned collection of operations on such partition sets.

**Definition 3.1** (set of weighted partitions). Recall that $\Pi(U)$ denotes the set of all partitions of some set $U$. A *set of weighted partitions* is a set $\mathcal{A} \subseteq \Pi(U) \times \mathbb{N}$, i.e., a family of pairs, each consisting of a partition of $U$ and a non-negative integer weight. We say that $\mathcal{A}$ is *unweighted* if $(p, w) \in \mathcal{A}$ implies that $w = 0$.

The operators naturally apply to connectivity problems by allowing, e.g., gluing of connected components (i.e., different sets in a partition), or joining of two partial solutions by taking the meet operation $\sqcap$ on the respective partitions. We will later see that if the recurrence only uses these operators, then the naive algorithm evaluating the recurrence can be improved beyond the typical $2^{\Omega(\mathtt{tw} \cdot \log \mathtt{tw})}$ that comes from the high number of different possible partial solutions.

For notational ease, we let $\mathtt{rmc}(\mathcal{A})$ denote the set obtained by removing non-minimal weight copies, i.e.,
$$\mathtt{rmc}(\mathcal{A}) = \left\{(p, w) \in \mathcal{A} \,\middle|\, \nexists (p, w') \in \mathcal{A} \wedge w' < w \right\}.$$

**Definition 3.2** (operators on weighted partitions). Let $U$ be a set and $\mathcal{A} \subseteq \Pi(U) \times \mathbb{N}$.

- **Union.** For $\mathcal{B} \subseteq \Pi(U) \times \mathbb{N}$, define $\mathcal{A} \uplus \mathcal{B} = \mathtt{rmc}(\mathcal{A} \cup \mathcal{B})$. *Combine two sets of weighted partitions and discard dominated partitions.*

- **Insert.** For $X \cap U = \emptyset$, define $\mathtt{ins}(X, \mathcal{A}) = \{(p_{\uparrow U \cup X}, w) \,|\, (p, w) \in \mathcal{A}\}$. *Insert additional elements into $U$ and add them as singletons in each partition.*

- **Shift.** For $w' \in \mathbb{N}$ define $\mathtt{shft}(w', \mathcal{A}) = \{(p, w + w') \,|\, (p, w) \in \mathcal{A}\}$. *Increase the weight of each partition by $w'$.*

- **Glue.** For $u, v$, let $\hat{U} = U \cup \{u, v\}$ and define $\mathtt{glue}(uv, \mathcal{A}) \subseteq \Pi(\hat{U}) \times \mathbb{N}$ as
$$\mathtt{glue}(uv, \mathcal{A}) = \mathtt{rmc}(\left\{(\hat{U}[uv] \sqcap p_{\uparrow \hat{U}}, w) \,\middle|\, (p, w) \in \mathcal{A}\right\}).$$

  Also, if $\omega : \hat{U} \times \hat{U} \to \mathbb{N}$, let $\mathtt{glue}_\omega(uv, \mathcal{A}) = \mathtt{shft}(\omega(u, v), \mathtt{glue}(uv, \mathcal{A}))$. *In each partition combine the sets containing $u$ and $v$ into one; add $u$ and $v$ to the base set if needed.*

- **Project.** For $X \subseteq U$ let $\overline{X} = U \setminus X$, and define $\mathtt{proj}(X, \mathcal{A}) \subseteq \Pi(\overline{X}) \times \mathbb{N}$ as
$$\mathtt{proj}(X, \mathcal{A}) = \mathtt{rmc}(\left\{(p_{\downarrow \overline{X}}, w) \,\middle|\, (p, w) \in \mathcal{A} \wedge \forall e \in X : \exists e' \in \overline{X} : p \sqsubseteq U[ee']\right\}).$$

  *Remove all elements of $X$ from each partition, but discard partitions where this would reduce the number of blocks/sets.*

- **Join.** For $\mathcal{B} \subseteq \Pi(U') \times \mathbb{N}$ let $\hat{U} = U \cup U'$ and define $\mathtt{join}(\mathcal{A}, \mathcal{B}) \subseteq \Pi(\hat{U}) \times \mathbb{N}$ as
$$\mathtt{join}(\mathcal{A}, \mathcal{B}) = \mathtt{rmc}(\left\{(p_{\uparrow \hat{U}} \sqcap q_{\uparrow \hat{U}}, w_1 + w_2) \,\middle|\, (p, w_1) \in \mathcal{A} \wedge (q, w_2) \in \mathcal{B}\right\}).$$

  *Extend all partitions to the same base set. For each pair of partitions return the outcome of the meet operation $\sqcap$, with weight equal to the sum of the weights.*

Regarding the definition, note the role of $U[ab]$: Using $p \sqcap U[ab]$ we obtain a partition that is the same as $p$ except that the sets containing $a$ and $b$ are now merged into one (if they were one set already then nothing happens). When we use $p \sqsubseteq U[ab]$ then this is true if $a$ and $b$ are in the same set in the partition $p$ (but the set can be larger than just $\{a, b\}$).

Note also that $\texttt{ins}(X, \mathcal{A})$ and $\texttt{shft}(,a)$ re the only operators that do not require $\texttt{rmc}()$ in its definition; all other operators may create weighted partitions that consist of the same partition but with different weights (we want to keep only the cheaper one).

Straightforward implementation gives the following:

**Proposition 3.3.** *Each of the operations union, shift, insert, glue and project can be performed in $S|U|^{\mathcal{O}(1)}$ time where $S$ is the size of the input of the operation. Given $\mathcal{A}, \mathcal{B}$, $\texttt{join}(\mathcal{A}, \mathcal{B})$ can be computed in time $|\mathcal{A}| \cdot |\mathcal{B}| \cdot |U|^{\mathcal{O}(1)}$.*

## 3.2 Representing collections of weighted partitions

The key idea for getting a faster dynamic programming algorithm is to follow the naive DP, but to consider only small representative sets of weighted partitions instead of all weighted partitions that would be considered by the naive DP. Intuitively, a representative (sub)set of partial solutions should allow us to always extend to an optimal solution provided that one of the complete set of partial solutions extends to it. Let us define this formally.

**Definition 3.4** (Representation). Given a set of weighted partitions $\mathcal{A} \subseteq \Pi(U) \times \mathbb{N}$ and a partition $q \in \Pi(U)$, define

$$\texttt{opt}(q, \mathcal{A}) = \min \{w \mid (p, w) \in \mathcal{A} \land p \sqcap q = \{U\}\}.$$

For another set of weighted partitions $\mathcal{A}' \subseteq \Pi(U) \times \mathbb{N}$, we say that $\mathcal{A}'$ *represents* $\mathcal{A}$ if for all $q \in \Pi(U)$ it holds that $\texttt{opt}(q, \mathcal{A}') = \texttt{opt}(q, \mathcal{A})$.

Note that the definition of representation is symmetric, i.e., if $\mathcal{A}'$ represents $\mathcal{A}$ then $\mathcal{A}$ also represents $\mathcal{A}'$. However, we will only be interested in the special case where $\mathcal{A}' \subseteq \mathcal{A}$ and where we have a size guarantee for finding a small such subset $\mathcal{A}'$.

**Definition 3.5** (Preserving representation). A function $f : 2^{\Pi(U) \times \mathbb{N}} \times Z \to 2^{\Pi(U') \times \mathbb{N}}$ is said to *preserve representation* if for every $\mathcal{A}, \mathcal{A}' \subseteq \Pi(U) \times \mathbb{N}$ and $z \in Z$ it holds that if $\mathcal{A}'$ represents $\mathcal{A}$ then $f(\mathcal{A}', z)$ represents $f(\mathcal{A}, z)$. (Note that $Z$ stands for any combination of further inputs.)

Completing the required tools, the following lemma and theorem establish that the operations needed for the DP preserve representation, and that we can always find a reasonably small representative set of weighted partitions. The proofs are deferred to Section 3.6.

**Lemma 3.6.** *The union, insert, shift, glue, project, and join operations from Definition 3.2 preserve representation.*

**Theorem 3.7** ($\texttt{reduce}$). *There exists an algorithm $\texttt{reduce}$ that given set of weighted partitions $\mathcal{A} \subseteq \Pi(U) \times \mathbb{N}$, outputs in $|\mathcal{A}|2^{(\omega-1)|U|}|U|^{\mathcal{O}(1)}$ time a set of weighted partitions $\mathcal{A}' \subseteq \mathcal{A}$ such that $\mathcal{A}'$ represents $\mathcal{A}$ and $|\mathcal{A}'| \leq 2^{|U|}$, where $\omega$ denotes the matrix multiplication exponent.*

It was recently shown that $\omega < 2.3727$ [47].

## 3.3 Application to STEINER TREE.

In this section we show how to solve the STEINER TREE problem via a dynamic programming formulation that requires only the operators introduced in Section 3.3.

> STEINER TREE
> **Input:** A graph $G = (V, E)$ weight function $\omega\colon E \to \mathbb{N} \setminus \{0\}$, a terminal set $K \subseteq V$ and a nice tree decomposition $\mathbb{T}$ of $G$ of width $\mathtt{tw}$.
> **Question:** The minimum of $\omega(X)$ over all subsets $X \subseteq E$ of $G$ such that $G[X]$ is connected and $K \subseteq V(G[X])$.

Of course, since the weights are positive, in an optimal solution $X$ indeed induces a tree as the problem name suggests. We will start by restating the recurrence used by the folklore algorithm for solving this problem using the introduced terminology on weighted partitions. Generally, we will denote the (to be computed) tables as $A_x(\cdot)$ whereas $\mathcal{E}_x(\cdot)$ stands for a set of partial solutions; in both cases $x$ denotes the current bag. For a bag $x$, and $\mathbf{s} \in \{0,1\}^{B_x}$ define

$$A_x(\mathbf{s}) = \left\{ \left( p, \min_{X \in \mathcal{E}_x(p,\mathbf{s})} \omega(X) \right) \middle| p \in \Pi(s^{-1}(1)) \wedge \mathcal{E}_x(p,\mathbf{s}) \neq \emptyset \right\}$$

$$\mathcal{E}_x(p, \mathbf{s}) = \Big\{ X \subseteq E_x \Big| \forall v \in B_x : v \in V(G[X]) \vee v \in K \to s(v) = 1$$
$$\wedge \forall v_1, v_2 \in s^{-1}(1) : v_1 v_2 \text{ are in same block in } p \leftrightarrow v_1, v_2 \text{ connected in } G[X]$$
$$\wedge \, \#\mathtt{blocks}(p) = \mathtt{cc}(G[X]) \Big\}.$$

Note that $s\colon B_x \to \{0,1\}$ expresses which vertices we chose for the Steiner tree (namely those with $s(v) = 1$). Intuitively, $\mathcal{E}_x(p, \mathbf{s})$ is the set of partial solutions having (a subset of) $s^{-1}(1)$ as incident vertices in $B_x$ and connecting the vertices of $s^{-1}(1)$ according to $p$. Furthermore, the definition ensures that all connected components spanned by the edges of any partial solution contain at least one vertex of the current bag, and all terminals are contained in some connected component. Thus if we pick the tree decomposition $\mathbb{T}$ such that the root is the forget node, say $x$, for some terminal, say $v$, then we can check at the (single) child $y$ of $x$ the entry $A_y(\mathbf{s})$ where $s(v) = 1$ (there are no other vertices in this bag): This allows only the partition $p = \{\{v\}\}$, and by definition it must correspond to a minimum weight set of edges that spans a single connected component that contains all terminals.

We proceed with the recurrence for $A_x(\mathbf{s})$ which is used by the folklore dynamic programming algorithm. In order to simplify the notation, let $v$ denote the vertex introduced and contained in an introduce bag, and let $y, z$ denote the left and right children of $x$ in $\mathbb{T}$, if present. We distinguish on the type of bag in $\mathbb{T}$:

**Leaf bag $x$:**
$$A_x(\emptyset) = \{(\emptyset, 0)\}$$

Indeed, $\mathcal{E}_x(p, \mathbf{s})$ only contains the empty set; it connects nothing and has weight 0.

**Introduce vertex $v$ bag $x$ with child $y$:**

$$A_x(\mathbf{s}) = \begin{cases} \mathtt{ins}(\{v\}, A_y(\mathbf{s}_{|B_y})), & \text{if } s(v) = 1 \\ A_y(\mathbf{s}_{|B_y}), & \text{if } s(v) = 0 \wedge v \notin K \\ \emptyset, & \text{if } s(v) = 0 \wedge v \in K \end{cases}$$

Using $v$ for a Steiner tree corresponds to $s(v) = 1$, and we insert $v$ as a singleton into each partition (as there are no edges incident with $v$ yet). If $v$ is a terminal, then not inserting $v$ is not feasible. Else, if $v \notin K$, and we do not use $v$ then we get the same partial solutions as the previous bag.

**Forget vertex $v$ bag $x$ with child $y$:**

$$A_x(\mathbf{s}) = A_y(\mathbf{s}[v \to 0]) \uplus \texttt{proj}(v, A_y(\mathbf{s}[v \to 1]))$$

We combine partial solutions that either include $v$ or exclude $v$. If $v$ is included, we ensure it was connected to other vertices by removing it with the project operation (if it was a singleton then the corresponding partition is effectively eliminated).

**Introduce edge $e = uv$ bag $x$ with child $y$:**

$$A_x(\mathbf{s}) = \begin{cases} A_y(\mathbf{s}) & \text{if } s(u) = 0 \vee s(v) = 0 \\ A_y(\mathbf{s}) \uplus \texttt{glue}_\omega(uv, A_y(\mathbf{s})), & \text{otherwise.} \end{cases}$$

To be able to include the edge we need by the definition of $\mathcal{E}_x(p, \mathbf{s})$ that $s(u) = s(v) = 1$. If we include the edge, we account for the weight $\omega(u, v)$ and update the connectivity $p$ by connecting $u$ and $v$ with the glue operation.

**Join bag $x$ with children $y$ and $z$:**

$$A_x(\mathbf{s}) = \texttt{join}(A_y(\mathbf{s}), A_z(\mathbf{s})).$$

We know that every partial solution represented in $A_x(\mathbf{s})$ can be obtained from partial solutions represented by $A_y(\mathbf{s})$ and $A_z(\mathbf{s})$. To combine two partial solutions from $A_y(\mathbf{s})$ and $A_z(\mathbf{s})$, we have to sum the weights of the used edges and join the connectivity, and this is exactly what the join operation does.

**Theorem 3.8.** *There exist algorithms that given a graph $G$ solve* STEINER TREE *in time $n(1 + 2^\omega)^{\texttt{pw}}\texttt{pw}^{\mathcal{O}(1)}$ time if a path decomposition of width $\texttt{pw}$ of $G$ is given, and in time $n(1 + 2^{\omega+1})^{\texttt{tw}}\texttt{tw}^{\mathcal{O}(1)}$ time if a tree decomposition of width $\texttt{tw}$ of $G$ is given.*

*Proof.* The algorithm is the following: use the above dynamic programming formulation as discussed to compute $A_r$ (where $r$ is the child of the root, as discussed), but after evaluation of every entry $A_x$, use Theorem 3.7 to obtain and store $A'_x = \texttt{reduce}(A_x)$ rather than $A_x$. Since $A_x = \texttt{reduce}(\mathcal{A})$ represents $\mathcal{A}$ and the recurrence uses only the operators defined in Definition 3.2 which all preserve representation by Lemma 3.6, we have as invariant that for every $x \in \mathbb{T}$ the entry $A'_x$ stored for $A_x$ represents $A_x$ by Lemma 3.6. In particular, the stored value $A'_r(\mathbf{s})$ represents $A_r(\mathbf{s})$ and hence we can safely read off the answer to the problem from this stored value as done from $A_r(\mathbf{s})$ in the folklore dynamic programming algorithm. Let us focus on the time analysis: since for all operations from Definition 3.2 we can apply Proposition 3.3, and for obtaining the tree/path decomposition with the required properties we can use Proposition 2.2 the bottleneck clearly is the `reduce` algorithm.

If we are given a path decomposition, it can easily be seen that the intermediate sets of weighted partitions are always of size at most $2^{|s^{-1}(1)|+1}$, and hence the time needed for computing $\texttt{reduce}(A_x(\mathbf{s}))$ for any bag $x$ can be upper bounded by $\sum_{i_0+i_1=|B_x|} \binom{|B_x|}{i_1} 1^{i_0} 2^{\omega i_1} \texttt{pw}^{\mathcal{O}(1)} = (1+2^\omega)^{\texttt{pw}}\texttt{pw}^{\mathcal{O}(1)}$.

If we are given a tree decomposition, we need to consider the time required to compute $A'_x(\mathbf{s})$ were $x$ is a join bag. Then the size of the intermediate sets of weighted partitions is easily upper bounded by $4^{|s^{-1}(1)|}$, and hence the time needed for computing $\texttt{reduce}(A_x(\mathbf{s}))$ can be upper bounded by $\sum_{i_0+i_1=|B_x|} \binom{|B_x|}{i_1} 1^{i_0} 2^{(\omega+1)i_1} \texttt{tw}^{\mathcal{O}(1)} = (1+2^{\omega+1})^{\texttt{tw}}\texttt{tw}^{\mathcal{O}(1)}$. □

## 3.4 Application to TRAVELING SALESMAN.

In this section we give an algorithm for TRAVELING SALESMAN by expressing a corresponding dynamic programming using the operators from Section 3.1.

> TRAVELING SALESMAN
> **Input:** A graph $G = (V, E)$, weight function $\omega \colon E \to \mathbb{N}$ and tree decomposition $\mathbb{T}$ of $G$ of width $\mathtt{tw}$.
> **Question:** The minimum of $\omega(X)$ over all Hamiltonian cycles $X \subseteq E$.

First, for technical convenience, we guess an edge $v_1 v_n \in X$ that has to be included in the Hamiltonian cycle. Note that this can be done with only $\mathtt{tw}$ guesses since we can guess an edge adjacent to a vertex of minimum degree which is easily seen to be at most $\mathtt{tw}$. Now, using Proposition 2.2, we turn the tree decomposition $\mathbb{T}$ into a nice tree decomposition that is rooted at the introduce edge bag for $\{v_1, v_n\}$; slightly abusing notation we refer to the latter as $\mathbb{T}$. After accounting for the weight of $\{v_1, v_n\}$, this reduces the problem to $\mathtt{tw}$ instances of finding a minimum weight Hamiltonian path between two fixed vertices.

For a bag $x$ and $\mathbf{s} \in \{0, 1, 2\}^{B_x}$ define

$$A_x(\mathbf{s}) = \left\{ \left(M, \min_{X \in \mathcal{E}_x(M, \mathbf{s})} \omega(X)\right) \;\middle|\; M \in \Pi_2(s^{-1}(1)) \wedge \mathcal{E}_x(M, \mathbf{s}) \neq \emptyset \right\}$$

$$\mathcal{E}_x(M, \mathbf{s}) = \big\{ X \subseteq E_x : v \in B_x \to \deg_X(v) = s(v)$$
$$\wedge\, v \in V_x \setminus B_x \to \deg_X(v) = 2$$
$$\wedge\, \{u, v\} \in M \to u \text{ and } v \text{ are connected in } G_x[X]$$
$$\wedge\, G_x[X] \text{ contains no cycles}\big\}.$$

In this case, $s \colon B_x \to \{0, 1, 2\}$ encodes the degree of vertices in $B_x$ in the corresponding partial solution (the degree bound and exclusion of cycles implies that these are collections of paths). Our partitions $M \in \Pi_2(s^{-1}(1))$ store the pairing of degree-1 vertices induced by connecting paths. Again naively implemented this would give $2^{\Omega(\mathtt{tw} \log \mathtt{tw})}$ partial solutions, but our concept of representative sets gives the desired single-exponential runtime.

Consider the table entry $A_z(\mathbf{s})$ where $z$ is the root of $\mathbb{T}$, $s(v_1) = s(v_n) = 1$ and for $v \neq v_1, v_n$ we have $s(v) = 2$. It is easy to see that this table entry is empty if there is no Hamiltonian path from $v_1$ to $v_n$ and otherwise it is $(\{\{v_1, v_n\}\}, w)$ where $w$ is the weight of the minimum Hamiltonian path. Hence, to solve the problem it suffices to compute $A_z(\mathbf{s})$.

**Leaf bag $x$:** We have $B_x = \emptyset$. For ease of notation we permit a single table entry for the empty partition of $B_x$ into vertices labeled 0, 1, and 2:

$$A_x(\emptyset) = \{(\emptyset, 0)\}$$

**Introduce vertex $v$ bag $x$ with child $y$:** We have $B_x = B_y \cup \{v\}$. For all $\mathbf{s} \in \{0, 1, 2\}^{B_x}$ we compute $A_x(\mathbf{s})$ as follows

$$A_x(\mathbf{s}) = \begin{cases} A_y(\mathbf{s}_{|B_y}) & \text{if } s(v) = 0, \\ \emptyset & \text{otherwise.} \end{cases}$$

Since $v$ is just introduced it cannot have neighbors in $V_x$ and there is no change in connectivity.

**Forget vertex $v$ bag $x$ with child $y$:** We have $B_y = B_x \cup \{v\}$. For all $\mathbf{s} \in \{0, 1, 2\}^{B_x}$ we compute $A_x(\mathbf{s})$ as follows

$$A_x(\mathbf{s}) = A_y(\mathbf{s}[v \to 2]).$$

The vertex $v$ cannot have neighbors outside $V_x$ so it must have its two neighbors in $V_x$. Note that $\mathbf{s}[v \to 2] \in \{0,1,2\}^{B_y}$, with $s(v) = 2$.

**Introduce edge** $e = uv$ **bag** $x$ **with child** $y$: We have $B_x = B_y$. For all $\mathbf{s} \in \{0,1,2\}^{B_x}$ we compute $A_x(\mathbf{s})$ as follows

$$A_x(\mathbf{s}) = A_y(\mathbf{s}) \uplus \begin{cases} \emptyset, & \text{if } s(u) = 0 \vee s(v) = 0, \\ \texttt{glue}_\omega(uv, A_y(\mathbf{s}[u, v \to 0])) & \text{if } s(u) = s(v) = 1, \\ \texttt{proj}(\{v\}, \texttt{glue}_\omega(uv, A_y(\mathbf{s}[u \to 0, v \to 1]))) & \text{if } s(u) = 1 \wedge s(v) = 2, \\ \texttt{proj}(\{u\}, \texttt{glue}_\omega(uv, A_y(\mathbf{s}[u \to 1, v \to 0]))) & \text{if } s(u) = 2 \wedge s(v) = 1, \\ \texttt{proj}(\{u,v\}, \texttt{glue}_\omega(uv, A_y(\mathbf{s}[u, v \to 1]))) & \text{otherwise.} \end{cases}$$

In all cases, an optimal solution may simply make no use of the edge $e$; hence we copy the set of matchings from the child accordingly (this gives the term $A_y(\mathbf{s})$). If $s(u)$ or $s(v)$ is 0, the edge cannot have been used. When considering partial solutions using $e$, we have to use the entry of $A_y$ where the degree of $u$ and $v$ is decreased by 1 to account for the edge. In the case that $s(u)$ and $s(v)$ are 1 we can simply add the edge to the matching. In the case that $s(u) = 1$ and $s(v) = 2$ (or the symmetric opposite), we have to glue $e$ to $M$, meaning that in the new matching $u$ is matched with the vertex, say $w$, that $v$ was previously matched with; here glue first creates a set $\{u,v,w\}$ and project shrinks it to $\{u,w\}$ (since $v$ now has degree two and its specific connectivity needs not be traced). In the case that $s(u)$ and $s(v)$ both are 2, including $e$ in a matching $M$ in $A_y(s)$ gives the matching $M \setminus \{au, bv\} \cup \{ab\}$ for some $a, b$. We have to ensure that $\{a,b\} \neq \{u,v\}$ which is done by projecting (if $\{a,b\} = \{u,v\}$ then projection eliminates the partition).

**Join bag** $x$ **with children** $y$ **and** $z$: We have $B_x = B_y = B_z$ and compute $A_x(\mathbf{s})$ for all $\mathbf{s} \in \{0,1,2\}^{B_x}$ as follows

$$A_x(\mathbf{s}) = \biguplus_{\mathbf{l}+\mathbf{r}=\mathbf{s}} \texttt{proj}(s^{-1}(2) \setminus (l^{-1}(2) \cup r^{-1}(2)), \texttt{join}(A_y(\mathbf{l}), A_z(\mathbf{r})))$$

Note here that $\mathbf{l}, \mathbf{r}, \mathbf{s} \in \{0,1,2\}^{B_x}$ hence the summation is vector summation. Since we combine two characteristics of partial solutions into a new one, the degrees of the left and right partial solution ($\mathbf{l}$ and $\mathbf{r}$) have to sum up to the degrees of the new one ($\mathbf{s}$). Two characteristics $(M_1, w_1) \in A_y(\mathbf{l})$ and $(M_2, w_2) \in A_z(\mathbf{r})$ combine to a characteristic of $A_x(\mathbf{s})$ if and only if $M_1 \cup M_2$ is acyclic which is equivalent to saying that all vertices in $s^{-1}(2)$ are connected to vertices in $s^{-1}(1)$ in the resulting partition $M_1 \sqcap M_2$, and the latter is ensured by the project operation.

Using the tools from Subsection 3.2, we obtain the following result. However, it should be noted that we give an improvement of this in Subsection 3.6; this uses a non-trivial result from [17].

**Theorem 3.9.** *There exist algorithms that given a graph $G$ solve* TRAVELING SALESMAN *in time $n(2 + 2^\omega)^{\texttt{pw}} \texttt{pw}^{\mathcal{O}(1)}$ time if a path decomposition of width $\texttt{pw}$ of $G$ is given, and in time $n(7 + 2^{(\omega+1)})^{\texttt{tw}} \texttt{tw}^{\mathcal{O}(1)}$ time if a tree decomposition of width $\texttt{tw}$ of $G$ is given.*

*Proof.* The algorithm is the following: use the above dynamic programming formulation from Subsection 3.4 to compute $A_r$, but after evaluation of every entry $A_x$, use Theorem 3.7 to store $A'_x = \texttt{reduce}(A_x)$ rather than $A_x$. Since $A'_x = \texttt{reduce}(\mathcal{A})$ represents $\mathcal{A}$ and the recurrence uses only the operators defined in Definition 3.2 which all preserve representation by Lemma 3.6, we have as invariant that for every $x \in \mathbb{T}$ the entry $A'_x$ represents $A_x$. In particular, the stored value $A'_z(\mathbf{s})$ represents $A_z(\mathbf{s})$ where $z$ is the root of $\mathbb{T}$ and hence we can safely read off the answer to the problem as done in the naive dynamic programming algorithm from $A_z(\mathbf{s})$).

Let us focus on the time analysis: since for all operations from Definition 3.2 we can apply Proposition 3.3, and for obtaining the tree/path decomposition with the required properties we can use Proposition 2.2 the bottleneck clearly is the `reduce` algorithm. Also, note that the guessing of the edge $v_1v_n$ gives overhead of at most `pw` or `tw` so this will be subsumed by the last term of the claimed running time.

In the first algorithm that is given a path decomposition, note we can assume $x$ is either a leaf, introduce vertex, forget vertex or introduce edge bag. In this case the size of the intermediate sets of weighted partitions is by our invariant at most $2^{|s^{-1}(1)|}$. Then the time needed to perform $\texttt{reduce}(A_x(\mathbf{s}))$ is at most $2^{\omega|s^{-1}(1)|}\texttt{pw}^{\mathcal{O}(1)}$ and the time to compute $A_x(\mathbf{s})$ for every $\{0,1,2\}^{B_x}$ is

$$\sum_{\mathbf{s}\in\{0,1,2\}^{B_x}} 2^{\omega|s^{-1}(1)|}\texttt{pw}^{\mathcal{O}(1)} \leq \sum_{i_0+i_1+i_2=|B_x|} \binom{|B_x|}{i_0,i_1,i_2} 1^{i_0} 1^{i_2} 2^{\omega i_1}\texttt{pw}^{\mathcal{O}(1)}$$
$$= (2+2^\omega)^{|B_x|}\texttt{pw}^{\mathcal{O}(1)},$$

where the last equality is due to the multinomial theorem. Hence $A'_x$ can be computed for every $x \in \mathbb{T}$ in the claimed time bound.

For tree decompositions, note that if $x$ is a join bag, and we denote $\hat{A}$ for the intermediate values in the computation of $A'(x)$ then

$$|\hat{A}_x(\mathbf{s})| \leq \sum_{\mathbf{l}+\mathbf{r}=\mathbf{s}} 2^{|l^{-1}(1)|}2^{|r^{-1}(1)|} = \prod_{i=1}^{|B_x|}\sum_{l+r=s_i} 2^{[l=1]+[r=1]} = 4^{|s^{-1}(1)|}6^{|s^{-1}(2)|}.$$

where in the first equality we expand into independent products over all coordinates of the vectors and use the following simple observation in the second equality

$$\sum_{l_i+r_i=s_i} 2^{[l_i=1]+[r_i=1]} = \begin{cases} 1 & \text{if } s_i = 0 \\ 4 & \text{if } s_i = 1 \\ 6 & \text{if } s_i = 2. \end{cases} \quad (1)$$

Using this, it can also be seen that the reduce operation to obtain $A'_x(\mathbf{s})$ is performed in time bounded by $4^{|s^{-1}(1)|}6^{|s^{-1}(2)|}2^{(\omega-1)|s^{-1}(1)|}|B_x|^{\mathcal{O}(1)}$. Then the total time needed to evaluate $A_x(\mathbf{s})$ for every $\mathbf{s} \in \{0,1,2\}^{B_x}$ is bounded by

$$\sum_{i_0+i_1+i_2=|B_x|} \binom{|B_x|}{i_0,i_1,i_2} 1^{i_0} 2^{(\omega+1)i_1} 6^{i_2} = n(7+2^{(\omega+1)})^{|B_x|},$$

by the multinomial theorem and the claim follows. □

### 3.5 Application to FEEDBACK VERTEX SET.

As a third application of the rank based approach we show how to formulate and solve FEEDBACK VERTEX SET using a dynamic programming formulation based on the operators introduced in Section 3.1.

> FEEDBACK VERTEX SET
> **Input:** An undirected graph $G = (V,E)$, an integer $k$ and a nice tree decomposition $\mathbb{T}$ of $G$ of width `tw`.
> **Question:** Find a set $Y \subseteq V$ such that $|Y| \leq k$ and $G[V \setminus Y]$ is a forest.

Using the operators for weighted partitions for solving FEEDBACK VERTEX SET requires a slight reformulation of the problem. The reason is that the operators are designed to maximize connectivity, whereas FEEDBACK VERTEX SET requires that we avoid creating cycles in introduce edge nodes of the tree decomposition. (It can be seen that certain more directly applicable operators do not preserve representation, e.g., selection of all partitions in which two vertices are connected, but we omit a detailed discussion at this point.)

The idea for the reformulation is to seek an induced subgraph that is a tree on at least $i = |V| - k$ vertices; this is equivalent to requiring maximum connectivity using only $i-1$ edges. Of course, this requires that the connected components created by the deletion of a feedback vertex set can be connected into a single tree. To this end, we make the following changes: We modify $G$, by adding a special universal vertex $v_0$ to it, i.e., it is adjacent to all vertices of $G$. Denote $E_0$ for the set of edges incident to $v_0$. Then we ask for a pair $(Y, Y_0)$ such that $Y \subseteq V \setminus \{v_0\}$, $|Y| \leq k$, $Y_0$ is a subset of $E_0$, and the graph $(V \setminus Y, E[V \setminus Y] \cup Y_0)$ is a tree. (By $E[X]$ we mean all edges in $E$ with both endpoints in $X$.) This is clearly equivalent to the original problem statement since $G[V \setminus (Y \cup \{v_0\})]$ can be extended to a tree in this way if and only if it is a forest. We emphasize that the tree must contain all edges between all selected original vertices and may contain any edges incident on $v_0$ (this in particular affects the possibilities and introduce edge bags). Note also that we can modify the tree decomposition $\mathbb{T}$ accordingly by adding $v_0$ to all bags and making it nice again.

For a bag $x$, integers $i, j$ and $\mathbf{s} \in \{0,1\}^{B_x}$ define

$$A_x(\mathbf{s}, i, j) = \{(p, 0) \mid p \in \Pi(s^{-1}(1)) \land \mathcal{E}_x(p, \mathbf{s}, i, j) \neq \emptyset\}$$
$$\mathcal{E}_x(p, \mathbf{s}, i, j) = \{(X, X_0) \in 2^{V_x} \times 2^{E_0 \cap E_x} \mid |X| = i \land |E_x[X \setminus \{v_0\}] \cup X_0| = j$$
$$\land X \cap B_x = s^{-1}(1) \land v_0 \in B_x \to s(v_0) = 1$$
$$\land \forall u \in X \setminus B_x \exists u' \in s^{-1}(1) : u, u' \text{ connected in } (X, E_x[X \setminus \{v_0\}] \cup X_0)$$
$$\land \forall v_1, v_2 \in s^{-1}(1) : v_1 v_2 \text{ are in same block in } p \leftrightarrow v_1, v_2 \text{ are connected}$$
$$\text{in } (X, E_x[X \setminus \{v_0\}] \cup X_0)\}.$$

In words, $(p, 0) \in A_x(s, i, j)$ indicates that there exists a subset $\{v_0\} \cap V_x \subseteq X \subseteq V_x$ with $X \cap B_x = s^{-1}(1)$ and a subset $X_0 \subseteq E_x \cap E_0$ such that in the graph $(X, E_x[X \setminus \{v_0\}] \cup X_0)$ we have $i$ vertices, $j$ edges, no connected component fully contained in $V_x \setminus B_x$, and that the elements of $s^{-1}(1)$ are connected according to the partition $p$. It follows that the given instance of FEEDBACK VERTEX SET is a YES-instance if and only if for some $i \geq |V \cup \{v_0\}| - k = |V| - (k-1)$ we have that $A_z(\emptyset, i, i-1)$ is non-empty: For the forward direction, we can take a solution and extend it to a tree of the required type using the incident edges of $v_0$. For the backward direction, we have that $(X, E_x[X \setminus \{v_0\}] \cup X_0)$ contains $i$ vertices and $i - 1$ edges and that it is connected; hence it is a tree and $V \setminus X$ is a feedback vertex set.

In order to simplify the notation, let $v$ denote the vertex introduced and contained in an introduce bag, and let $y, z$ denote the left and right children of $x$ in $\mathbb{T}$, if present. As usual, undefined table entries are assumed to be empty for notational convenience. We distinguish on the type of bag in $\mathbb{T}$:

**Leaf bag $x$:**
$$A_x(\emptyset, 0, 0) = \{(\emptyset, 0)\}$$

Clearly, if $i = j = 0$ then $\mathcal{E}_x(\emptyset, i, j) = \{(\emptyset, \emptyset)\}$ and it is empty otherwise.

**Introduce vertex $v$ bag $x$ with child $y$:**

$$A_x(\mathbf{s}, i, j) = \begin{cases} \emptyset & \text{if } v = v_0 \wedge s(v) = 0, \\ \texttt{ins}(\{v\}, A_y(\mathbf{s}_{|B_y}, i-1, j)), & \text{if } s(v) = 1, \\ A_y(\mathbf{s}_{|B_y}, i, j), & \text{otherwise.} \end{cases}$$

If $v = v_0$ we require $s(v) = 1$ by definition. Otherwise, if $s(v) = 1$, we account for its inclusion in $X$: We extend solutions where the number of vertices is $i-1$ (effectively, this increases the number of vertices, as intended).

**Forget vertex $v$ bag $x$ with child $y$:**

$$A_x(\mathbf{s}, i, j) = A_y(\mathbf{s}[v \to 0], i, j) \uplus \texttt{proj}(v, A_y(\mathbf{s}[v \to 1], i, j)).$$

We combine solutions from the child bag where $v$ can be included or excluded in the intended tree (since the current bag does not specify $s(v)$). For previous solutions with $s(v) = 1$, we remove $v$ from the partitions using the project operator; this eliminates partial solutions that have $v$ as a singleton in the partition since that would create a separate component that can no longer be connected (deviating from building a single tree).

**Introduce edge $e = uv$ bag $x$ with child $y$:**

$$A_x(\mathbf{s}, i, j) = \begin{cases} A_y(\mathbf{s}, i, j) \uplus \texttt{glue}(v_0 v, A_y(\mathbf{s}, i, j-1)) & \text{if } u = v_0 \wedge s(v) = 1, \\ \texttt{glue}(uv, A_y(\mathbf{s}, i, j-1)) & \text{if } s(u) = 1 \wedge s(v) = 1, \\ A_y(\mathbf{s}, i, j) & \text{otherwise.} \end{cases}$$

If $u = v_0$ (or, by symmetry, $v = v_0$) we can choose to insert the edge $v_0 v$, and account for it by extending previous solutions with $j-1$ edges. If $s(u) = s(v) = 1$ we have to include the edge $uv$ in $E_x[X]$ and again build upon solutions with $j-1$ edges.

**Join bag $x$ with children $y$ and $z$:**

$$A_x(\mathbf{s}, i, j) = \biguplus_{\substack{i_1 + i_2 = i + s^{-1}(1) \\ j_1 + j_2 = j}} \texttt{join}(A_y(\mathbf{s}, i_1, j_1), A_z(\mathbf{s}, i_2, j_2)).$$

We first compensate for vertices accounted for twice in the $i$ counter since they were introduced in both subtrees of $\mathbb{T}$ by subtracting their number; these are exactly the vertices in $s^{-1}(1) \subseteq B_x$ (they are selected for the tree and by basic tree decomposition properties they must be contained in $B_x$ if they are in both subtrees). Then we simply add the number of edges of both subsolutions, and their connectivity is automatically joined by the join operator.

We obtain the following theorem, whose proof is analogous to the proof of Theorem 3.8 and therefore omitted. However, let us explain why we can still obtain a time bound that is linear in $n$, despite the two additional table dimensions $i$ and $j$ of range $\mathcal{O}(n)$. The trick here is to see that we do not need to fully evaluate the table, but that smaller ranges for $i$ and $j$ suffice. First, let us see that $i \in \{j+1, \ldots, j + \texttt{tw} + 1\}$: If $j \geq i$ then the partial solution must already contain a cycle (and every further added vertex needs at least one incident edge so we cannot reach $i^*$ vertices and $i^* - 1$ edges). If $i > j + \texttt{tw} + 1$ then the table is empty, since matching partial solutions would have more than $\texttt{tw} + 1$ connected components, which violates the condition that each component is connected to a vertex of the current bag. Second, to see that a small range of values for $i$ suffices at each bag, consider the following: If we have a nonempty table entry $A_x(\mathbf{s}, i, j)$, then any table entry $A_x(\mathbf{s}', i', j')$ with $i' < i - \mathtt{tw}$ is suboptimal (and hence not required): We can always add $B_x \setminus \{v_0\}$ to the implicit feedback vertex set of any solution in $A_x(\mathbf{s}, i, j)$ to get a better solution than $A_x(\mathbf{s}', i', j')$; it is less constrained since its connected components contain only the mandatory $v_0$ from $B_x$ and it has at least $i'$ vertices in connected components.

**Theorem 3.10.** *There exist algorithms that given a graph $G$ solve* FEEDBACK VERTEX SET *in time $n(1 + 2^\omega)^{\mathtt{pw}}\mathtt{pw}^{\mathcal{O}(1)}$ time if a path decomposition of width $\mathtt{pw}$ of $G$ is given, and in time $n(1 + 2^{\omega+1})^{\mathtt{tw}}\mathtt{tw}^{\mathcal{O}(1)}$ time if a tree decomposition of width $\mathtt{tw}$ of $G$ is given.*

## 3.6 Representing collections of weighted partitions based on rank

**Preserving representation.** We first need to prove Lemma 3.6. That is, that the operations union, insert, shift, glue, project, and join preserve representation. While this is an essential part of the rank based approach, the actual proofs are not particularly instructive to read and are postponed to Appendix B.3.

**Finding small representative subsets.** The remainder of this subsection is devoted to the proof of Theorem 3.7. The key idea is as follows: To find optimal partial solutions in a set $\mathcal{A}$ of weighted partitions, one checks for certain partitions $q$ what is the minimum weight $w$ such that $\mathcal{A}$ contains some pair $(p, w)$ such that $p \sqcap q$ gives the unit partition. For intuition let us ignore the weights for the moment and only look at the partitions in $\mathcal{A}$. Very roughly we need to find a subset of those partitions that can complete any partition $q$ to the unit partition $\{U\}$, assuming that some partition in $\mathcal{A}$ does that too (we only need to represent what $\mathcal{A}$ can do). It can be seen that any set cover of partitions $p$ such that for any $q$ at least one of them gives $p \sqcap q = \{U\}$ suffices. It turns out that taking a subset of partitions that form a basis in a certain matrix can play the same role, and is much easier to handle (both in proof and for finding it algorithmically). Given that, it is not hard to get the additional property that the representative subset always matches the correct weight that the original set $\mathcal{A}$ would provide; this corresponds essentially to a basis of minimum weight.

Our matrix simply states for all pairs $p$ and $q$ of partitions whether or not the meet operation applied to them gives the unit partition. The crucial part, of course, is to show that it has low rank, in order to guarantee a small basis. The matrix is formally defined as follows; for notational ease let us fix $U = [t]$, and let us shorthand $(V_1, V_2)$ for a partition $\{V_1, V_2\} \in \Pi(U)$.

**Definition 3.11.** Define $\mathcal{M} \in \mathbb{Z}_2^{\Pi(U) \times \Pi(U)}$ by

$$\mathcal{M}[p, q] = \begin{cases} 1 & p \sqcap q = \{U\}, \\ 0 & \text{else.} \end{cases}$$

The idea for getting a good rank bound for this matrix is to consider a simple class of partitions, namely cuts into only two sets (where the second set may in fact be empty). The subsequent lemma then shows that in arithmetic modulo two, the matrix $\mathcal{M}$ can be written as the product of two cutmatrices $\mathcal{C}$, which are defined as follows.

**Definition 3.12.** Define $\mathtt{cuts}(t) := \{(V_1, V_2) \mid V_1 \,\dot\cup\, V_2 = U \wedge 1 \in V_1\}$, where 1 stands for an arbitrary but fixed element of $U$. Define $\mathcal{C} \in \mathbb{Z}_2^{\Pi(U) \times \mathtt{cuts}(t)}$ by $\mathcal{C}[p, (V_1, V_2)] = [(V_1, V_2) \sqsubseteq p]$.

Intuitively, the matrix $\mathcal{C}$ represents which pairs of partitions and cuts are consistent, i.e., the partition needs to be a refinement of the cut. It is crucial that one side of the cut is fixed to contain a particular vertex since that allows the following trick via counting modulo two: For each pair of partitions, the product $\mathcal{CC}^T$ can be seen to count the number of partitions that are consistent with both $p$ and $q$. If $p \sqcap q = \{U\}$ then there is only one such partition. Otherwise, if $p \sqcap q$ is a partition with at least two sets then there is an even number of cuts that are consistent (to see this, pair the cuts by including respectively excluding any set that does not include the special element $1 \in V_1$). The formal proof is postponed to Appendix B.1.

**Lemma 3.13.** *It holds that $\mathcal{M} \equiv \mathcal{CC}^T$.*

From the factorization we pretty immediately get the following lemma that encapsulates the idea behind finding small representative subsets. It shows that linear dependence allows us to discard one of the corresponding partitions (e.g., the one of highest weight).

**Lemma 3.14.** *Let $X \subseteq \Pi(U)$ and $q, r \in \Pi(U)$ such that $\mathcal{C}[q, \cdot] \equiv \sum_{p \in X} \mathcal{C}[p, \cdot]$ and $r \sqcap q = \{U\}$. Then, there exists $p \in X$ such that $r \sqcap p = \{U\}$.*

*Proof.* We have that

$$\mathcal{M}[q, \cdot] \equiv \mathcal{C}[q, \cdot]\mathcal{C}^T \equiv (\sum_{p \in X} \mathcal{C}[p, \cdot])\mathcal{C}^T \equiv \sum_{p \in X} \mathcal{M}[p, \cdot].$$

Where the equivalences respectively follow by Lemma 3.13, assumption, and Lemma 3.13 combined with linearity of matrix multiplication. In particular, $\mathcal{M}[q, r] \equiv \sum_{p \in X} \mathcal{M}[p, r]$. Thus, if $\mathcal{M}[q, r] = 1$ then $\mathcal{M}[p, r] = 1$ for some $p \in X$ and the lemma follows. □

At this point any naive algorithm for finding a lightest basis would suffice to complete our algorithm, e.g., this is straightforward via Gaussian elimination when the rows are ordered by weight. However, aiming for a faster algorithm, the following lemma finds the required basis faster via matrix multiplication; the proof is postponed to Appendix B.2.

**Lemma 3.15.** *There is an algorithm that, given a $n \times m$ matrix with $m \leq n$ and with entries from the field $\mathbb{F}_2$ and weights $\omega : [n] \to \mathbb{N}$, finds a basis $X \subseteq [n]$ of the row space minimizing $\omega(X)$ in $\mathcal{O}(nm^{\omega-1})$.*

Now we can wrap up and prove Theorem 3.7.

*Proof of Theorem 3.7.* The algorithm `reduce` is as follows:

---
**Algorithm** `reduce(𝒜)`
1: let $\mathcal{A} = \{(p_1, w_1), \ldots, (p_\ell, w_\ell)\}$ and $A = \{p_1, \ldots, p_\ell\}$.
2: **if** $\ell \leq 2^{|U|}$ **then return** $\mathcal{A}$
3: find a basis $X \subseteq A$ of the row space of $\mathcal{C}[A, \cdot]$ minimizing $\sum_{p_i \in X} w_i$ using Lemma 3.15.
4: **return** $\{(p, w) \mid (p, w) \in \mathcal{A} \land p \in X\}$.

---

Clearly, $\mathcal{A}' = \text{reduce}(\mathcal{A}) \subseteq \mathcal{A}$. Since the number of columns of $\mathcal{C}$ equals $2^{|U|-1}$, $X$ being a basis for the row space guarantees that $|\mathcal{A}'| \leq 2^{|U|}$. The running time is clearly dominated by the call to the algorithm of Lemma 3.15 since any entry of $\mathcal{C}$ can be computed in $|U|^{\mathcal{O}(1)}$ time. Then, since $\mathcal{C}[A, \cdot]$ is a $|\mathcal{A}| \times 2^{t-2}$ the claimed running time follows.

It remains to argue that $\mathcal{A}'$ represents $\mathcal{A}$. Suppose for a contradiction that this is not the case. Since $\mathcal{A}' \subseteq \mathcal{A}$ we have for every $q$ that $\text{opt}(q, \mathcal{A}) \leq \text{opt}(q, \mathcal{A}')$. Thus our assumption implies that $\text{opt}(q, \mathcal{A}) < \text{opt}(q, \mathcal{A}')$, that is, for some $q$ there is $(p, w) \in \mathcal{A}$ and $q$ such that $p \sqcap q = \{U\}$ and $w < \texttt{opt}(q, \mathcal{A}')$. Since $X$ is a basis of the row space of $\mathcal{C}[A, \cdot]$ we know that the row $\mathcal{C}[p, \cdot]$ is a linear combination of a set of rows from $\mathcal{C}[X, \cdot]$. Let $Y \subseteq X$ be the set of indices of these rows. By Lemma 3.14, we know that there exists $p_i \in Y$ such that $q \sqcap p_i = \{U\}$; Hence, since $w < \texttt{opt}(q, \mathcal{A}')$, it must be that $w_i > w$. But we have that $(X \setminus \{p_i\}) \cup \{p\}$ also is a basis of the row space of $\mathcal{C}[A, \cdot]$ since $\mathcal{C}[p_i, \cdot]$ is a linear combination of the rows indexed by $(Y \setminus \{p_i\}) \cup \{p\}$. Since $w_i > w$, this contradicts that $X$ minimizes $\sum_{p_i \in X} w_i$. □

## 3.7 Improvements and limitations of the approach

The main ingredient for the proof of Theorem 3.7 is Lemma 3.14 which in turn follows directly from the factorization $\mathcal{M} = \mathcal{C}\mathcal{C}^T$ (Lemma 3.13). Thus by establishing a better factorization, i.e., with smaller inner dimension, using matrices whose entries are computable in polynomial time, then we would immediately get improved algorithmic results.

It can be seen that such an improvement is not possible: Assume that $|U|$ is odd and fix an element $u_0 \in U$. The submatrix of $\mathcal{M}$ with rows and columns indexed by $\{U[X] \mid u_0 \in X \wedge |X| = (|U|+1)/2\}$ is easily seen to be an identity matrix, since the meet of two of such partitions gives the unit partition if and only if they are constructed from $X_1$ and $X_2$ with $X_1 = U \setminus X_2 \cup \{u_0\}$. Since the inner dimension is an upper bound on the rank this rules out relevant improvements to the factorization of $\mathcal{M}$, because the above construction shows that the rank is at least $2^{|U|}/|U|$.

However, if we restrict our attention to the submatrix of $\mathcal{M}$ corresponding only to perfect matchings (such as we use for the TSP algorithm) then the following improved factorization can be established.

**Theorem 3.16** ([17]). *Let $\mathcal{H}$ be the submatrix of $\mathcal{M}$ restricted to all matchings. Then $\mathcal{H}$ can be factorized into two matrices whose entries can be computed in time $|U|^{\mathcal{O}(1)}$, where the inner dimension of the factorization is $2^{|U|/2-1}$.*

Combining this result with the proof technique of Theorem 3.7, we obtain the following improvement:

**Corollary 3.17.** *There exists an algorithm $\texttt{reducematchings}$ that given set of weighted matchings $\mathcal{A} \subseteq \Pi_2(U) \times \mathbb{N}$, outputs in $|\mathcal{A}|2^{\frac{(\omega-1)}{2}|U|}|U|^{\mathcal{O}(1)}$ time a set of weighted matchings $\mathcal{A}' \subseteq \mathcal{A}$ such that $\mathcal{A}'$ represents $\mathcal{A}$ and $|\mathcal{A}'| \leq 2^{|U|/2}$, where $\omega$ denotes the matrix multiplication exponent.*

*Proof.* The proof is almost identical to the proof of Theorem 3.7: let $\mathcal{H} = \mathcal{C}'\mathcal{C}''$ be the matrix factorization from Theorem 3.16. Then by the same proof, Lemma 3.14 holds when we replace $\mathcal{C}$ with $\mathcal{C}'$ and restrict $X$ and $p$ and $r$ to consist of matchings. Then $\texttt{reducematchings}$ is obtained from $\texttt{reduce}$ by replacing $\mathcal{C}$ with $\mathcal{C}'$. The arguments for the correctness and running times of this algorithm are analogous to the arguments from the proof of Theorem 3.7. □

Thus, for problems whose dynamic programming formulation via our introduced set of operators requires only weighted perfect matchings, but not the generality of weighted partitions, we get somewhat faster algorithms; among the considered problems this is true for TRAVELING SALESMAN, and it seems unlikely for STEINER TREE or FEEDBACK VERTEX SET.

**Theorem 3.18.** *There exist algorithms that given a graph $G$ solve TRAVELING SALESMAN in time $n(2 + 2^{\omega/2})^{\texttt{pw}}\texttt{pw}^{\mathcal{O}(1)}$ time if a path decomposition of width $\texttt{pw}$ of $G$ is given, and in time $n(5 + 2^{(\omega+2)/2})^{\texttt{tw}}\texttt{tw}^{\mathcal{O}(1)}$ time if a tree decomposition of width $\texttt{tw}$ of $G$ is given.*

The proof is analogous to the proof of Theorem 3.9, the only difference being the use of $\texttt{reducematchings}$ instead of $\texttt{reduce}$; for completeness a proof is provided in Appendix B.4.

Combining the ideas from the proof of Theorem 3.18 with Lemma 1.1 and a trick from [18] we obtain the following corollary:

**Corollary 3.19.** *There is an algorithm solving* Traveling Salesman *on cubic graphs in* $1.2186^n n^{\mathcal{O}(1)}$ *time.*

The observation from [18] is that on cubic graphs, we know that the degree of a vertex in a subsolution cannot be 2 if at most 1 incident edges is introduced, an it cannot be 0 if at least 2 edges are introduced since the remaining edge is not enough to make the degree 2. Hence these states can be safely ignored.

Again the proof idea similar to proof of Theorem 3.18 we can also rather easily obtained the following result:

**Theorem 3.20.** *There exist algorithms that given a graph $G$ and integer $k$ finds a path of length $k$ in time $n(2 + 2^{\omega/2})^{\mathtt{pw}}(k + \mathtt{pw})^{\mathcal{O}(1)}$ time if a path decomposition of width $\mathtt{pw}$ of $G$ is given, and in time $n(5 + 2^{(\omega+2)/2})^{\mathtt{tw}}(k + \mathtt{tw})^{\mathcal{O}(1)}$ time if a tree decomposition of width $\mathtt{tw}$ of $G$ is given.*

The idea is to slightly modify the table entries from Subsection 3.4: In the definition of $\mathcal{E}_x$ we remove the requirement $v \in V_x \setminus B_x \to \deg_X(v) = 2$, and hence the formula in the forget bag has to be altered to $A_x(\mathbf{s}) = A_y(\mathbf{s})$. Then we can set all weights to 1 and now we are up to maximizing $\omega(X)$. It is easy to see that our approach can be altered to deal with maximization problems and hence this gives the required result since involved weight being more than $k \cdot \mathtt{tw}$ already guarantees that we have a $k$-path.

# 4 Determinant approach

In this section we will present the determinant approach that can be used to solve counting versions of connectivity problems on graphs of small treewidth. Throughout this section, we will assume a graph $G$ along with a path/tree decomposition $\mathbb{T}$ of $G$ of with $\mathtt{pw}$ or $\mathtt{tw}$ is given.

Let $A$ be an incidence matrix of an orientation of $G$, that is $A = (a_{i,j})$ is a matrix with $n$ rows and $m$ columns. Each row of $A$ is indexed with a vertex and each column of $A$ is indexed with an edge. The entry $a_{v,e}$ is defined to be 0 if $v \notin e$; $-1$ if $e = uv$ and $u < v$; or 1 if $e = uv$ and $u > v$. We assume, that all the vertices are ordered with respect to the post-ordering of their forget nodes in the tree, that is vertices forgotten in a left subtree are smaller than vertices forgotten in the right subtree, and a vertex forgotten in a bag $x$ is smaller than a vertex forgotten in a bag which is an ancestor of $x$. Similarly we order edges according to the post-ordering of the introduce edge nodes in the tree decomposition.

Let $v_1$ be an arbitrary fixed vertex and let $F$ be the matrix $A$ with the row corresponding to $v_1$ removed. For a subset $S \subseteq E$ let $F_S$ be the matrix with $n - 1$ rows and $|S|$ columns, whose columns are those of $F$ with indices in $S$. The following folklore lemma is used in the proof of the Matrix Tree Theorem (see for example [1, Page 203] where our matrix is denoted by $N$)

**Lemma 4.1.** *Let $S \subseteq E$ be a subset of size $n - 1$. If $(V, S)$ is a tree, then $|\det(F_S)| = 1$ and $\det(F_S) = 0$ otherwise.*

In this section we are up to compute the number of connected edgesets $X$ such that $X \in \mathcal{F}$ where $\mathcal{F}$ is some implicitly defined set family. Our main idea is to use Lemma 4.1 to reduce this task to computing the quantity $\sum_{X \in \mathcal{F}} \det(F_X)^2$ instead, and to ensure that if $X \in \mathcal{F}$ is connected, then it is a tree.

For two (not necessarily disjoint) subsets $V_1, V_2$ of an ordered set let us define $\mathrm{inv}(V_1, V_2) = |\{(u, v) : u \in V_1, v \in V_2, u > v\}|$. If $X, Y$ are ordered set, recall that for a permutation $f : X \overset{1-1}{\to} Y$ we have that the sign equals $\mathrm{sgn}(f) = (-1)^{|\{(e_1, e_2) : e_1, e_2 \in S \wedge e_1 < e_2 \wedge f(e_1) > f(e_2)\}|}$.

The following proposition will be useful:

**Proposition 4.2.** *Let $X_l, X_r \subseteq V$ and $Y_l, Y_r \subseteq E$ such that $X_l \cap X_r = \emptyset$ and $Y_l \cap Y_r = \emptyset$ and for every $e_1 \in Y_l$ and $e_2 \in Y_r$ we have that $e_1 < e_2$. Suppose $f_l : Y_l \xrightarrow{1-1} X_l$ and $f_r : Y_r \xrightarrow{1-1} X_r$. Denote $f = f_l \cup f_r$, that is, $f(v) = f_l(v)$ if $v \in Y_l$ and $f(v) = f_r(v)$ if $v \in Y_r$. Then it holds that $\mathrm{sgn}(f) = \mathrm{sgn}(f_1)\mathrm{sgn}(f_2) \cdot (-1)^{\mathrm{inv}(X_l, X_r)}$.*

To see that the proposition is true, note that from the definition of sgn, the pairs $e_1, e_2$ with $e_1, e_2 \in Y_1$ or $e_1, e_2 \in Y_2$ are already accounted for in the part $\mathrm{sgn}(f_1)\mathrm{sgn}(f_2)$ so it remains to show that

$$|\{(e_1, e_2) : e_1 \in Y_1, e_2 \in Y_2 \wedge e_1 < e_2 \wedge f(e_1) > f(e_2)\}|$$

indeed equals $\mathrm{inv}(X_l, X_r)$, but this is easy to see since we have by assumption that $e_1 < e_2$.

## 4.1 Counting Hamiltonian cycles

For our first application to counting Hamiltonian cycles, we derive the following formula which expresses the number of Hamiltonian cycles of a graph:

$$\sum_{\substack{X \subseteq E \\ \forall v \in V \, \deg_X(v) = 2}} [X \text{ is a Hamiltonian cycle}]$$

$\{$a 2-regular graph has $n$ subtrees on $n-1$ edges if it is connected and 0 otherwise$\}$

$$= \frac{1}{n} \cdot \sum_{\substack{X \subseteq E \\ \forall v \in V \, \deg_X(v) = 2}} \sum_{S \subseteq X, |S| = n-1} [(V, S) \text{ is a tree}]$$

$\{$Lemma 4.1$\}$

$$= \frac{1}{n} \cdot \sum_{\substack{X \subseteq E \\ \forall v \in V \, \deg_X(v) = 2}} \sum_{S \subseteq X, |S| = n-1} \det(F_S)^2.$$

By plugging the permutation definition of a determinant, we obtain the following expression for the number of Hamiltonian cycles of a graph:

$$\frac{1}{n} \cdot \sum_{\substack{X \subseteq E \\ \forall v \in V \, \deg_X(v) = 2}} \sum_{S \subseteq X, |S| = n-1} \Big( \sum_{f : S \xrightarrow{1-1} V \setminus \{v_1\}} \mathrm{sgn}(f) \prod_{e \in S} a_{f(e),e} \Big)^2$$

$$= \frac{1}{n} \cdot \sum_{\substack{X \subseteq E \\ \forall v \in V \, \deg_X(v) = 2}} \sum_{S \subseteq X} \sum_{f_1, f_2 : S \xrightarrow{1-1} V \setminus \{v_1\}} \mathrm{sgn}(f_1)\mathrm{sgn}(f_2) \prod_{e \in S} a_{f_1(e),e} a_{f_2(e),e},$$

Note that in the last equality we dropped the assumption $|S| = n - 1$, as it follows from the fact that $f_1$ (and $f_2$) is a bijection.

Our goal is to compute the formula by dynamic programming over some nice tree decomposition $\mathbb{T}$. To this end, let us define a notion of "partial sum" of the above formula, that we will store in our dynamic programming table entries. For every bag $x \in \mathbb{T}$, $s_{\deg} \in \{0, 1, 2\}^{B_x}$, $s_1 \in \{0, 1\}^{B_x}$ and $s_2 \in \{0, 1\}^{B_x}$ define

$$A_x(s_{\deg}, s_1, s_2) = \sum_{\substack{X \subseteq E_x \\ \forall_{v \in (V_x \setminus B_x)} \deg_X(v)=2 \\ \forall_{v \in B_x} \deg_X(v)=s_{\deg}(v)}} \sum_{S \subseteq X} \sum_{\substack{f_1: S \overset{1-1}{\to} (V_x \setminus \{v_1\}) \setminus s_1^{-1}(0) \\ f_2: S \overset{1-1}{\to} (V_x \setminus \{v_1\}) \setminus s_2^{-1}(0)}} \text{sgn}(f_1)\text{sgn}(f_2) \prod_{e \in S} a_{f_1(e),e} a_{f_2(e),e}$$
(2)

Intuitively in $s_{\deg}$ we store the degrees of vertices of $B_x$ in $G[X]$, whereas $s_1$ (and $s_2$) specify whether a vertex of $B_x$ was already used as a value of the bijection $f_1$ (and $f_2$).

**Leaf bag $x$:**
$$A_x(\emptyset, \emptyset, \emptyset) = 1$$

**Introduce vertex $v$ bag $x$ with child $y$:** for $s_{\deg} \in \{0,1,2\}^{B_x}$ and $s_1, s_2 \in \{0,1\}^{B_x}$

$$A_x(s_{\deg}, s_1, s_2) = \begin{cases} A_y(s_{\deg}|_{B_y}, s_1|_{B_y}, s_2|_{B_y}) & \text{if } s_{\deg}(v) = s_1(v) = s_2(v) = 0 \\ 0 & \text{otherwise.} \end{cases}$$

Since the vertex $v$ is not incident to any edge from $E_x$, we have that $\deg_X = 0$ and hence all summands in which $f_1$ or $f_2$ map an edge to $v$ will vanish.

**Forget vertex $v$ bag $x$ with child $y$:** for $s_{\deg} \in \{0,1,2\}^{B_x}$ and $s_1, s_2 \in \{0,1\}^{B_x}$

$$A_x(s_{\deg}, s_1, s_2) = \begin{cases} A_y(s_{\deg}[v \to 2], s_1[v \to 1], s_2[v \to 1]) & \text{if } v \neq v_1 \\ A_y(s_{\deg}[v \to 2], s_1[v \to 0], s_2[v \to 0]) & \text{otherwise} \end{cases}$$

Since $v$ is moved out of $B_x$ we need to make sure $\deg_X(v) = 2$, and since no edges incident to $v$ will be introduced anymore we require $s_1(v) = s_2(v) = 1$, unless $v = v_1$: Since it is not used as a value for bijections $f_1, f_2$, we need to handle this vertex separately.

**Introduce edge $e = uv$ bag $x$ with child $y$:** Here we have three cases. Either the edge $e$ is not contained in the set $X$, or it is a part of $X \setminus S$, or finally it is a part of $S$. In the last case we need to choose the values $f_1(e)$ and $f_2(e)$ and account for their contribution to $\text{sgn}(f_1)\text{sgn}(f_2)$, but we can restrict ourselves to values of $(s_1^{-1}(1) \setminus \{v_1\}) \cap e$ and $(s_2^{-1}(1) \setminus \{v_1\}) \cap e$ respectively, because otherwise either we do not obtain a bijection or the whole summand will disappear since $a_{v',e} = 0$ for $v' \notin e$.

If $s_{\deg}(u), s_{\deg}(v) \geq 1$, then we set

$$\begin{aligned} A_x(s_{\deg}, s_1, s_2) =& A_y(s_{\deg}, s_1, s_2) + A_y(s'_{\deg}, s_1, s_2) \\ &+ \sum_{\substack{u' \in (s_1^{-1}(1) \setminus \{v_1\}) \cap e \\ v' \in (s_2^{-1}(1) \setminus \{v_1\}) \cap e}} A_y(s'_{\deg}, s_1[u' \to 0], s_2[v' \to 0]) \cdot a_{u',e} \cdot a_{v',e} \\ &\cdot (-1)^{\text{inv}(s_1^{-1}(1), u') + \text{inv}(s_2^{-1}(1), v')}, \end{aligned}$$

where $s'_{\deg}$ is the function $s_{\deg}$ with the values for $u$ and $v$ decreased by one. Observe that on the first two mentioned cases are accounted for on the first line. For the third case, we need to account for the new number of inversions contributing to $\text{sgn}(f_1)\text{sgn}(f_2)$. Note that vertices from $V_x \setminus B_x$ cannot participate in such an inversion since the vertex and edge are both smaller than $v'$ and $e$ due to the ordering.

To this end, we can use Proposition 4.2 with $Y_r = \{e\}$ and $X_r = \{v'\}$ since $e$ is the largest edge in $E_x$, and it follows that the factor $(-1)^{\text{inv}(s_1^{-1}(1),u')+\text{inv}(s_2^{-1}(1),v')}$ indeed accounts for the new inversions.

Finally, when $s_{\deg}(u) = 0$ or $s_{\deg}(v) = 0$, then we set
$$A_x(s_{\deg}, s_1, s_2) = A_y(s_{\deg}, s_1, s_2)$$

**Join bag $x$ with children $y$ and $z$:**

$$A_x(s_{\deg}, s_1, s_2) = \sum_{\substack{s_{\deg,y}+s_{\deg,z}=s_{\deg} \\ s_{1,y}+s_{1,z}=s_1 \\ s_{2,y}+s_{2,z}=s_2}} A_y(s_{\deg,y}, s_{1,y}, s_{2,y}) \cdot A_z(s_{\deg,z}, s_{1,z}, s_{2,z})$$
$$\cdot (-1)^{\text{inv}(s_{1,y}^{-1}(1), s_{1,z}^{-1}(1)) + \text{inv}(s_{2,y}^{-1}(1), s_{2,z}^{-1}(1))}$$

In the above formula when adding two functions we denote coordinate-wise addition. To see that we correctly account for the new number of inversions, first note that vertices from $V_x \setminus B_x$ cannot participate in such an inversion since the vertex and incident edge are both smaller than the vertices appearing in $B_x$ and their incident edges $e$ due to the ordering. Then by Proposition 4.2 we correctly account for the new inversions since all edges from $E_y$ are smaller than $E_z$ due to the post-ordering of edges.

**Theorem 4.3.** *There exist algorithms that given a graph $G$ solve $\#$ HAMILTONIAN CYCLE in $\tilde{\mathcal{O}}(6^{\texttt{pw}}\texttt{pw}^{\mathcal{O}(1)}n^2)$ time if a path decomposition of width $\texttt{pw}$ is given, and in time $\tilde{\mathcal{O}}(15^{\texttt{tw}}\texttt{tw}^{\mathcal{O}(1)}n^2)$ time if a tree decomposition of width $\texttt{tw}$ is given.*

*Proof.* Observe that when all the table entries of a tree decomposition rooted at an empty bag $z$ are computed, the number of Hamiltonian cycles is equal to $A_z(\emptyset, \emptyset, \emptyset)/n$. Moreover each of the $\texttt{tw}^{\mathcal{O}(1)}n$ nodes involves operations on integers of bitlength $O(n \log n)$, since all the table entries have absolute value at most $n^{O(n)}$.

In the described dynamic programming we have $3 \cdot 2 \cdot 2 = 12$ states per vertex, however observe that when $s_{\deg}(v) = 0$, then all non-zero summands of the partial sum (2) satisfy $s_1(v) = s_2(v) = 0$, since otherwise the function $f_1$ (or $f_2$) assigns the value of $v$ to an edge that was not incident to $v$. Similarly when $v \neq v_1$ and $s_{\deg}(v) = 2$, then it is enough to store the states only with $s_1(v) = s_2(v) = 1$, because if $v$ was not yet assigned as a value by $f_1$ (or $f_2$), then since we can not add to $X$ any more edges incident to $v$, this table entry will not be used as in the forget node we require that $s_1(v) = s_2(v) = 1$. Hence for each vertex $v \neq v_1$ there are only 6 triples $(s_{\deg}(v), s_1(v), s_2(v))$, i.e. $(0,0,0)$, $(1,0,0)$, $(1,0,1)$, $(1,1,0)$, $(1,1,1)$ and $(2,1,1)$, which is enough to show the claimed running time for path decompositions.

In case of tree decompositions we need to analyze the time complexity needed to handle the summation in join nodes. Observe that there are exactly 15 pairs of triples, which describe states of vertices in the left and right subtree, according to the following table, where **X** denotes that this pair is not a valid pair of triples. Hence we can restrict ourselves to these triples when evaluating the summation. □

## 4.2 Counting Steiner Trees

In this section we show how to count the number of Steiner trees of a prescribed size. Let $v_1$ be an arbitrary fixed terminal from $K$. The number of Steiner trees with exactly $k$ edges is expressed by the following formula.

|   | (0,0,0) | (1,0,0) | (1,0,1) | (1,1,0) | (1,1,1) | (2,1,1) |
|---|---------|---------|---------|---------|---------|---------|
| (0,0,0) | (0,0,0) | (1,0,0) | (1,0,1) | (1,1,0) | (1,1,1) | (2,1,1) |
| (1,0,0) | (1,0,0) | **X** | **X** | **X** | (2,1,1) | **X** |
| (1,0,1) | (1,0,1) | **X** | **X** | (2,1,1) | **X** | **X** |
| (1,1,0) | (1,1,0) | **X** | (2,1,1) | **X** | **X** | **X** |
| (1,1,1) | (1,1,1) | (2,1,1) | **X** | **X** | **X** | **X** |
| (2,1,1) | (2,1,1) | **X** | **X** | **X** | **X** | **X** |

$$\sum_{X\subseteq E, |X|=k} [X \text{ is a Steiner tree}] = \sum_{Y\subseteq V, |Y|=k-1, K\subseteq Y} \sum_{X\subseteq E(Y,Y), |X|=|Y|-1} [(Y,X) \text{ is a tree}]$$

$$= \sum_{Y\subseteq V, |Y|=k-1, K\subseteq Y} \sum_{X\subseteq E(Y,Y), |X|=|Y|-1} \det(F_{Y,X})^2$$

where $F_{Y,X}$ is the submatrix of $F$ with rows in $Y \setminus \{v_1\}$ and columns in $X$. Again, by plugging the permutation definition of a determinant we obtain:

$$\sum_{Y\subseteq V, |Y|=k-1, K\subseteq Y} \sum_{X\subseteq E(Y,Y), |X|=|Y|-1} \sum_{f_1, f_2: X \overset{1-1}{\to} Y\setminus\{v_1\}} \text{sgn}(f_1)\text{sgn}(f_2) \prod_{e\in X} a_{f_1(e),e} a_{f_2(e),e},$$

where $\text{sgn}(f) = (-1)^{|\{(e_1,e_2): e_1,e_2 \in X \wedge e_1 < e_2 \wedge f(e_1) > f(e_2)\}|}$.

Let us define a "partial sum" of the above formula. For every bag $x \in \mathbb{T}$, $0 \leq i \leq |V_x|$, $s_Y \in \{0,1\}^{B_x}$, $s_1 \in \{0,1\}^{B_x}$ and $s_2 \in \{0,1\}^{B_x}$ define:

$$A_x(i, s_Y, s_1, s_2) = \sum_{\substack{Y\subseteq V_x \\ |Y|=i \\ (K\cap V_x)\subseteq Y \\ Y\cap B_x = s_Y^{-1}(1)}} \sum_{X\subseteq E(Y,Y)\cap E_x} \sum_{\substack{f_1: X \overset{1-1}{\to} Y\setminus\{v_1\}\setminus s_1^{-1}(0) \\ f_2: X \overset{1-1}{\to} Y\setminus\{v_1\}\setminus s_2^{-1}(0)}} \text{sgn}(f_1)\text{sgn}(f_2) \prod_{e\in X} a_{f_1(e),e} a_{f_2(e),e}.$$

Note that in the above definition when $s_Y(v) = 0$, then in order to have a non-zero table entry we need to have $s_1(v) = s_2(v) = 0$, since otherwise an edge of $X$ would be assigned (by $f_1$ or $f_2$) a vertex which is not its endpoint, which leaves only 5 reasonable states per vertex.

**Leaf bag $x$:**
$$A_x(0, \emptyset, \emptyset, \emptyset) = 1$$

**Introduce vertex $v$ bag $x$ with child $y$:** for $0 \leq i \leq |V_x|$, $s_Y \in \{0,1\}^{B_x}$ and $s_1, s_2 \in \{0,1\}^{B_x}$

$$A_x(i, s_Y, s_1, s_2) = \begin{cases} A_y(i, s_Y|_{B_y}, s_1|_{B_y}, s_2|_{B_y}) & \text{if } s_Y(v) = s_1(v) = s_2(v) = 0, v \notin K \\ A_y(i-1, s_Y|_{B_y}, s_1|_{B_y}, s_2|_{B_y}) & \text{if } s_Y(v) = 1, s_1(v) = s_2(v) = 0 \\ 0 & \text{otherwise.} \end{cases}$$

Note that the choice of including (or not) the vertex $v$ into $Y$ is described by $s_Y(v)$ and moreover if $v \in K$ then we have to include $v$ into $Y$.

**Forget vertex $v$ bag $x$ with child $y$:** for $0 \leq i \leq |V_x|$, $s_Y \in \{0,1\}^{B_x}$ and $s_1, s_2 \in \{0,1\}^{B_x}$

$$A_x(i, s_Y, s_1, s_2) = \begin{cases} A_y(i, s_Y[v \to 1], s_1[v \to 1], s_2[v \to 1]) & \text{if } v \in K, v \neq v_1 \\ A_y(i, s_Y[v \to 1], s_1[v \to 0], s_2[v \to 0]) & \text{if } v = v_1 \\ A_y(i, s_Y[v \to 1], s_1[v \to 1], s_2[v \to 1]) & \\ +A_y(i, s_Y[v \to 0], s_1[v \to 0], s_2[v \to 0]) & \text{otherwise} \end{cases}$$

**Introduce edge $e = uv$ bag $x$ with child $y$:**

$$A_x(i, s_Y, s_1, s_2) = A_y(i, s_Y, s_1, s_2)$$
$$+ \sum_{\substack{u' \in (s_1^{-1}(1) \setminus \{v_1\}) \cap e \\ v' \in (s_2^{-1}(1) \setminus \{v_1\}) \cap e}} A_y(i, s_Y, s_1[u' \to 0], s_2[v' \to 0]) \cdot a_{u',e} \cdot a_{v',e}$$
$$\cdot [s_Y(u) = s_Y(v) = 1](-1)^{\text{inv}(s_1^{-1}(1),u') + \text{inv}(s_2^{-1}(1),v')}$$

**Join bag $x$ with children $y$ and $z$:**

$$A_x(s_Y, s_1, s_2) = \sum_{\substack{s_{1,y} + s_{1,z} = s_1 \\ s_{2,y} + s_{2,z} = s_2}} A_y(s_Y, s_{1,y}, s_{2,y}) \cdot A_z(s_Y, s_{1,z}, s_{2,z})$$
$$\cdot (-1)^{\text{inv}(s_{1,y}^{-1}(1), s_{1,z}^{-1}(1)) + \text{inv}(s_{2,y}^{-1}(1), s_{2,z}^{-1}(1))}$$

Observe that the functions $s_Y$ is used in both children, to make the choice of taking a vertex to $Y$ or not consistent between the two subtrees.

**Theorem 4.4.** *There exist algorithms that given a graph $G$ counts the number of Steiner trees of size $i$ for each $1 \leq i \leq n-1$ in $\tilde{\mathcal{O}}(5^{\mathtt{pw}}\mathtt{pw}^{\mathcal{O}(1)}n^3)$ time if a path decomposition of width $\mathtt{pw}$ is given, and in time $\tilde{\mathcal{O}}(10^{\mathtt{tw}}\mathtt{tw}^{\mathcal{O}(1)}n^3)$ time if a tree decomposition of width $\mathtt{tw}$ is given.*

*Proof.* Observe that when all the table entries of a tree decomposition rooted at an empty bag $z$ are computed, the number of Steiner trees with exactly $i$ edges is equal to $A_z(i+1, \emptyset, \emptyset, \emptyset)$. Moreover each of the $tw^{\mathcal{O}(1)}n$ nodes involes operations on integers of bitlength $O(n \log n)$, since all the table entries have absolute value at most $n^{O(n)}$.

As we have already noted we can assume that for each vertex $v \neq v_1$ the triple $(s_Y(v), s_1(v), s_2(v))$ is of one of the following forms $(0,0,0)$, $(1,0,0)$, $(1,0,1)$, $(1,1,0)$, $(1,1,1)$, which is enough to prove the claimed running time for path decompositions.

In order to handle join nodes effectively, observe that there are exactly 10 pairs of triples, which describe states of vertices in the left and right subtree, according to the following table, where **X** denotes that this pair is not a valid pair of triples.

|           | $(0,0,0)$ | $(1,0,0)$ | $(1,0,1)$ | $(1,1,0)$ | $(1,1,1)$ |
|-----------|-----------|-----------|-----------|-----------|-----------|
| $(0,0,0)$ | $(0,0,0)$ | **X**     | **X**     | **X**     | **X**     |
| $(1,0,0)$ | **X**     | $(1,0,0)$ | $(1,0,1)$ | $(1,1,0)$ | $(1,1,1)$ |
| $(1,0,1)$ | **X**     | $(1,0,1)$ | **X**     | $(1,1,1)$ | **X**     |
| $(1,1,0)$ | **X**     | $(1,1,0)$ | $(1,1,1)$ | **X**     | **X**     |
| $(1,1,1)$ | **X**     | $(1,1,1)$ | **X**     | **X**     | **X**     |

□

## 4.3 Feedback Vertex Set

Note that to solve the FEEDBACK VERTEX SET problem it is enough to solve its dual, which will be easier to work with in the squared determinant framework.

---
MAXIMUM INDUCED FOREST
**Input:** An undirected graph $G$, an integer $k$ and a nice tree decomposition $\mathbb{T}$ of $G$ of width `tw`.
**Question:** Find a set $Y \subseteq V$ such that $|Y| \geq k$ and $G[Y]$ is a forest.

---

Without loss of generality we can assume that the graph $G$ contains an isolated vertex, which we denote as $v_1$. Since if such vertex is not present we can add it, increase $k$ by one and update the tree decomposition accordingly. Clearly $v_1$ is contained in any maximal induced forest. So far by using squared determinant we counted trees, but now we want to check whether there exists an induced forest of a prescribed size. To achieve this we define a new set of edges $E'$ containing all the edges between $v_1$ and $V \setminus \{v_1\}$ (note that $E' \cap E(G) = \emptyset$). Observe, that there exists an induced forest containing $v_1$ with $n_0$ vertices and $m_0$ edges (hence $n - m_0$ connected components) in $G$ iff there exists an induced *tree* in $G' = (V, E(G) \cup E')$ containing $v_1$ with $n_0$ vertices and $n - m_0 - 1$ edges of $E'$. Therefore it is enough to count induced trees in $G'$, while keeping track of the number of used edges from $E'$. Consequently it is enough to find, whether the following sum is positive

$$\sum_{\substack{Y \subseteq V, v_1 \in Y, |Y|=k}} \sum_{\substack{X \subseteq E(Y,Y) \cup E'(Y,Y) \\ E(Y,Y) \subseteq X \\ |X|=|Y|-1}} [(Y,X) \text{ is a tree }]$$

$$= \sum_{\substack{Y \subseteq V, v_1 \in Y, |Y|=k}} \sum_{\substack{X \subseteq E(Y,Y) \cup E'(Y,Y) \\ E(Y,Y) \subseteq X \\ |X|=|Y|-1}} \det(F'_{Y,X})^2$$

where $F'_{Y,X}$ is an orientation of an incidence matrix of $G$ with rows from $Y \setminus \{v_1\}$ and columns from $X$. By pluggin the permutation definition of the determinant we obtain the following

$$\sum_{\substack{Y \subseteq V, v_1 \in Y, |Y|=k}} \sum_{\substack{X \subseteq E(Y,Y) \cup E'(Y,Y) \\ E(Y,Y) \subseteq X}} \sum_{f_1,f_2: X \overset{1-1}{\to} Y \setminus \{v_1\}} \text{sgn}(f_1)\text{sgn}(f_2) \prod_{e \in X} a_{f_1(e),e} a_{f_2(e),e}$$

where $\text{sgn}(f) = (-1)^{|\{(e_1,e_2): e_1,e_2 \in X \land e_1 < e_2 \land f(e_1) > f(e_2)\}|}$. Note that we have removed the assumption that $|X| = |Y| - 1$ as this is enforced by the assumption that $f_1$ is a bijection.

Observe, that a tree decomposition of $G$ one can obtain a tree decomposition of $G'$ of width increase by at most one by ensuring that $v_1$ is included in every bag. Therefore we assume that we are given a nice tree decomposition $\mathbb{T}$ of $G'$. Let us define a "partial sum" of the above formula. For every bag $x \in \mathbb{T}$, $0 \leq i \leq |V_x|$, $s_Y \in \{0,1\}^{B_x}$, $s_1 \in \{0,1\}^{B_x}$ and $s_2 \in \{0,1\}^{B_x}$ define:

$$A_x(i, s_Y, s_1, s_2) = \sum_{\substack{Y \subseteq V_x \\ (\{v_1\} \cap V_x) \subseteq Y \\ |Y|=i \\ Y \cap B_x = s_Y^{-1}(1)}} \sum_{\substack{X \subseteq E_x \\ E_x \cap E(Y,Y) \subseteq X}} \sum_{\substack{f_1: X \overset{1-1}{\to} Y \setminus \{v_1\} \setminus s_1^{-1}(0) \\ f_2: X \overset{1-1}{\to} Y \setminus \{v_1\} \setminus s_2^{-1}(0)}} \text{sgn}(f_1)\text{sgn}(f_2) \prod_{e \in X} a_{f_1(e),e} a_{f_2(e),e}$$

The dynamic programming part of our algorithm is very similar to the one described for STEINER TREE.

**Leaf bag $x$:**
$$A_x(0, \emptyset, \emptyset, \emptyset) = 1$$

**Introduce vertex $v$ bag $x$ with child $y$:** for $0 \leq i \leq |V_x|$, $s_Y \in \{0,1\}^{B_x}$ and $s_1, s_2 \in \{0,1\}^{B_x}$

$$A_x(i, s_Y, s_1, s_2) = \begin{cases} A_y(i, s_Y|_{B_y}, s_1|_{B_y}, s_2|_{B_y}) & \text{if } s_Y(v) = s_1(v) = s_2(v) = 0, v \neq v_1 \\ A_y(i-1, s_Y|_{B_y}, s_1|_{B_y}, s_2|_{B_y}) & \text{if } s_Y(v) = 1, s_1(v) = s_2(v) = 0 \\ 0 & \text{otherwise.} \end{cases}$$

Note that $v_1$ has to be included into $Y$.

**Forget vertex $v$ bag $x$ with child $y$:** for $0 \leq i \leq |V_x|$, $s_Y \in \{0,1\}^{B_x}$ and $s_1, s_2 \in \{0,1\}^{B_x}$

$$A_x(i, s_Y, s_1, s_2) = \begin{cases} A_y(i, s_Y[v \to 1], s_1[v \to 0], s_2[v \to 0]) & \text{if } v = v_1 \\ A_y(i, s_Y[v \to 1], s_1[v \to 1], s_2[v \to 1]) & \\ + A_y(i, s_Y[v \to 0], s_1[v \to 0], s_2[v \to 0]) & \text{otherwise} \end{cases}$$

**Introduce edge $e = uv$ bag $x$ with child $y$:**

$$A_x(i, s_Y, s_1, s_2) = [e \in E' \vee s_Y(u) = 0 \vee s_Y(v) = 0] A_y(i, s_Y, s_1, s_2)$$
$$+ \sum_{\substack{u' \in (s_1^{-1}(1) \setminus \{v_1\}) \cap e \\ v' \in (s_2^{-1}(1) \setminus \{v_1\}) \cap e}} A_y(i, s_Y, s_1[u' \to 0], s_2[v' \to 0]) \cdot a_{u',e} \cdot a_{v',e}$$
$$\cdot [s_Y(u) = s_Y(v) = 1] (-1)^{\operatorname{inv}(s_1^{-1}(1), u') + \operatorname{inv}(s_2^{-1}(1), v')}$$

Observe that we have an option of not including $e$ into $X$ only if $e \in E'$ or $\min(s_Y(u), s_Y(v)) = 0$.

**Join bag $x$ with children $y$ and $z$:**

$$A_x(s_Y, s_1, s_2) = \sum_{\substack{s_{1,y} + s_{1,z} = s_1 \\ s_{2,y} + s_{2,z} = s_2}} A_y(s_Y, s_{1,y}, s_{2,y}) \cdot A_z(s_Y, s_{1,z}, s_{2,z})$$
$$\cdot (-1)^{\operatorname{inv}(s_{1,y}^{-1}(1), s_{1,z}^{-1}(1)) + \operatorname{inv}(s_{2,y}^{-1}(1), s_{2,z}^{-1}(1))}$$

Observe that the functions $s_Y$ is used in both children, to make the choice of taking a vertex to $Y$ or not consistent between the two subtrees.

**Theorem 4.5.** *There exist algorithms that given a graph $G$ solves the* FEEDBACK VERTEX SET *problem in $\tilde{\mathcal{O}}(5^{\texttt{pw}} \texttt{pw}^{\mathcal{O}(1)} n^3)$ time if a path decomposition of width $\texttt{pw}$ is given, and in time $\tilde{\mathcal{O}}(10^{\texttt{tw}} \texttt{tw}^{\mathcal{O}(1)} n^3)$ time if a tree decomposition of width $\texttt{tw}$ is given.*

*Proof.* Observe that when all the table entries of a tree decomposition rooted at an empty bag $z$ are computed, the there exists an induced forest in $G$ with $n_0$ vertices iff $A_z(n_0 + 1, \emptyset, \emptyset, \emptyset) > 0$. Moreover each of the $tw^{\mathcal{O}(1)} n$ nodes involes operations on integers of bitlength $O(n \log n)$, since all the table entries have absolute value at most $n^{O(n)}$.

Similarly as in the case of STEINER TREE we can assume that each vertex $v \neq v_1$ the triple $(s_Y(v), s_1(v), s_2(v))$ is of one of the following forms $(0,0,0)$, $(1,0,0)$, $(1,0,1)$, $(1,1,0)$, $(1,1,1)$, which is enough to prove the claimed running time for path decompositions.

In order to handle join nodes effectively, observe that there are exactly 10 pairs of triples, which describe states of vertices in the left and right subtree, according to the following table, where **X** denotes that this pair is not a valid pair of triples. The table is identical to the one used in the STEINER TREE problem. □

|           | $(0,0,0)$ | $(1,0,0)$ | $(1,0,1)$ | $(1,1,0)$ | $(1,1,1)$ |
|-----------|-----------|-----------|-----------|-----------|-----------|
| $(0,0,0)$ | $(0,0,0)$ | **X**     | **X**     | **X**     | **X**     |
| $(1,0,0)$ | **X**     | $(1,0,0)$ | $(1,0,1)$ | $(1,1,0)$ | $(1,1,1)$ |
| $(1,0,1)$ | **X**     | $(1,0,1)$ | **X**     | $(1,1,1)$ | **X**     |
| $(1,1,0)$ | **X**     | $(1,1,0)$ | $(1,1,1)$ | **X**     | **X**     |
| $(1,1,1)$ | **X**     | $(1,1,1)$ | **X**     | **X**     | **X**     |

# 5 Conclusions

In this paper, we have given deterministic algorithms for connectivity problems on graphs of small treewidth, with the running time only single exponential in the treewidth. We have given two different techniques. Each technique solves the standard unweighted versions, but for variants (counting, weighted versions, ...), sometimes only one of the techniques appears to be usable. Both techniques make novel use of classic linear algebra. As the treewidth and branchwidth of a graph differ by a constant factor, our algorithms also work for graphs of bounded branchwidth, and one can easily translate our algorithms to algorithms working on branch decompositions.

Our work improves upon the Cut&Count approach not only by having deterministic algorithms, but also (some of) our algorithms only use time that is linear in the number of vertices in the graph. However, the base of the exponent is somewhat larger, and an important question is whether this can be reduced. In the rank-based approach, this amounts to the following question: Can Algorithm `reduce` from the proof of Theorem 3.7 be implemented in linear time? It would also be interesting to have a linear time implementation of Algorithm `reducematchings` from Corollary 3.17.

We believe that our algorithms not only break a theoretical barrier, but may also be of practical use. A possible algorithm for e.g., Hamiltonian Circuit may be the following: run the standard DP on the tree decomposition, but as soon as we work with a table of size larger than $2^{\mathtt{tw}}$, we can use the rank-based approach and remove table entries by finding a base. Thus, we are guaranteed that we never have to process tables of size larger than $2^{\mathtt{tw}}$, at the cost of running a number of Gaussian elimination steps. It would be very interesting to perform an algorithm engineering study on this approach. It also may be interesting to see how this combines with a well known heuristic by Cook and Seymour for TSP [15], that uses finding a minimum length Hamiltonian Circuit in a graph with small branchwidth as a central step.

A final very intriguing question that our work implies is the following: Is the rank-based approach also usable in settings outside graphs of small treewidth? The rank-based approach gives a new twist to the dynamic programming approach, in the sense that we consider the "algebraic structure" of the partial certificates in a quite novel way. Thus, this suggests a study of this algebraic structure for dynamic programming algorithms for problems in other areas.

We would like to mention that uses of the rank based approach could also be found in work on rank-width boolean-width [30, 11], but in these cases the rank structure exploited is by more explicit assumption present in the input.

# Acknowledgements

We would like to thank Piotr Sankowski for pointing us to relevant literature of matrix-multiplication time algorithms. Moreover the second author thanks Marcin Pilipczuk and Łukasz Kowalik for helpful discussions at the early stage of the paper.

# References

[1] M. Aigner and G.M. Ziegler. Proofs from the book. *Berlin. Germany*, 2010.

[2] Noga Alon, Raphael Yuster, and Uri Zwick. Color-coding. *J. ACM*, 42(4):844–856, 1995.

[3] Stefan Arnborg and Andrzej Proskurowski. Linear time algorithms for NP-hard problems restricted to partial k-trees. *Discrete Applied Mathematics*, 23(1):11–24, 1989.

[4] Andreas Björklund. Determinant sums for undirected Hamiltonicity. In *FOCS*, pages 173–182. IEEE Computer Society, 2010.

[5] Andreas Björklund, Thore Husfeldt, Petteri Kaski, and Mikko Koivisto. Trimmed Möbius inversion and graphs of bounded degree. *Theory Comput. Syst.*, 47(3):637–654, 2010.

[6] Andreas Björklund, Thore Husfeldt, and Nina Taslaman. Shortest cycle through specified elements. In Yuval Rabani, editor, *SODA*, pages 1747–1753. SIAM, 2012.

[7] Hans Bodlaender. Treewidth: Structure and algorithms. In Giuseppe Prencipe and Shmuel Zaks, editors, *Structural Information and Communication Complexity*, volume 4474 of *Lecture Notes in Computer Science*, pages 11–25. Springer Berlin / Heidelberg, 2007.

[8] Hans L. Bodlaender. A linear-time algorithm for finding tree-decompositions of small treewidth. *SIAM J. Comput.*, 25(6):1305–1317, 1996.

[9] Hans L. Bodlaender and Arie M. C. A. Koster. Combinatorial Optimization on Graphs of Bounded Treewidth. *The Computer Journal*, 51(3):255–269, 2008.

[10] H.L. Bodlaender. On linear time minor tests with depth-first search. *Journal of Algorithms*, 14(1):1 – 23, 1993.

[11] Binh-Minh Bui-Xuan, Jan Arne Telle, and Martin Vatshelle. Boolean-width of graphs. *Theor. Comput. Sci.*, 412(39):5187–5204, 2011.

[12] P. Bürgisser, M. Clausen, and M.A. Shokrollahi. *Algebraic complexity theory*, volume 315. Springer, 1997.

[13] Yixin Cao, Jianer Chen, and Yang Liu. On feedback vertex set new measure and new structures. pages 93–104, 2010.

[14] J. Chen, J. Kneis, S. Lu, D. Mölle, S. Richter, P. Rossmanith, S. Sze, and F. Zhang. Randomized divide-and-conquer: Improved path, matching, and packing algorithms. *SIAM Journal on Computing*, 38(6):2526–2547, 2009.

[15] William Cook and Paul D. Seymour. Tour merging via branch-decomposition. *INFORMS Journal on Computing*, 15(3):233–248, 2003.

[16] Bruno Courcelle. The monadic second-order logic of graphs. i. recognizable sets of finite graphs. *Inf. Comput.*, 85(1):12–75, 1990.

[17] Marek Cygan, Stefan Kratsch, and Jesper Nederlof. Fast Hamiltonicity checking via bases of perfect matchings. 2012.

[18] Marek Cygan, Jesper Nederlof, Marcin Pilipczuk, Michal Pilipczuk, Johan M. M. van Rooij, and Jakub Onufry Wojtaszczyk. Solving connectivity problems parameterized by treewidth in single exponential time. In Rafail Ostrovsky, editor, *FOCS*, pages 150–159. IEEE, 2011.

[19] Erik D. Demaine, Mohammad Taghi Hajiaghayi, and Ken ichi Kawarabayashi. Algorithmic graph minor theory: Decomposition, approximation, and coloring. pages 637–646, 2005.

[20] Frederic Dorn, Fedor V. Fomin, and Dimitrios M. Thilikos. Catalan structures and dynamic programming in $H$-minor-free graphs. pages 631–640, 2008.

[21] Frederic Dorn, Eelko Penninkx, Hans L. Bodlaender, and Fedor V. Fomin. Efficient exact algorithms on planar graphs: Exploiting sphere cut decompositions. 58(3):790–810, 2010.

[22] David Eppstein. The traveling salesman problem for cubic graphs. 11(1):61–81, 2007.

[23] M. R. Fellows and M. A. Langston. On search decision and the efficiency of polynomial-time algorithms. In *Proceedings of the twenty-first annual ACM symposium on Theory of computing*, STOC '89, pages 501–512, New York, NY, USA, 1989. ACM.

[24] Fedor V. Fomin, Serge Gaspers, Saket Saurabh, and Alexey A. Stepanov. On two techniques of combining branching and treewidth. 54(2):181–207, 2009.

[25] Fedor V. Fomin and Dimitrios M. Thilikos. A simple and fast approach for solving problems on planar graphs. pages 56–67, 2004.

[26] Heidi Gebauer. On the number of Hamilton cycles in bounded degree graphs. pages 241–248, 2008.

[27] Mika Göös and Jukka Suomela. Locally checkable proofs. In Cyril Gavoille and Pierre Fraigniaud, editors, *PODC*, pages 159–168. ACM, 2011.

[28] Illya V. Hicks, Arie M. C. A. Koster, and Elif Kolotoğlu. Branch and tree decomposition techniques for discrete optimization. *Tutorials in Operations Research 2005*, pages 1–19, 2005.

[29] John E. Hopcroft, Rajeev Motwani, and Jeffrey D. Ullman. *Introduction to Automata Theory, Languages, and Computation*. 2nd edition, 2001.

[30] Sang il Oum and Paul Seymour. Approximating clique-width and branch-width. *Journal of Combinatorial Theory, Series B*, 96(4):514 – 528, 2006.

[31] Russell Impagliazzo, Ramamohan Paturi, and Francis Zane. Which problems have strongly exponential complexity? *J. Comput. Syst. Sci.*, 63(4):512–530, 2001.

[32] Kazuo Iwama and Takuya Nakashima. An improved exact algorithm for cubic graph TSP. pages 108–117, 2007.

[33] Jon Kleinberg and Éva Tardos. *Algorithm Design*. 2005.

[34] Ton Kloks. *Treewidth, Computations and Approximations*, volume 842 of *Lecture Notes in Computer Science*. Springer, 1994.

[35] Ioannis Koutis. Faster algebraic algorithms for path and packing problems. In Luca Aceto, Ivan Damgård, Leslie Ann Goldberg, Magnús M. Halldórsson, Anna Ingólfsdóttir, and Igor Walukiewicz, editors, *ICALP (1)*, volume 5125 of *Lecture Notes in Computer Science*, pages 575–586. Springer, 2008.

[36] Ioannis Koutis and Ryan Williams. Limits and applications of group algebras for parameterized problems. In Susanne Albers, Alberto Marchetti-Spaccamela, Yossi Matias, Sotiris E. Nikoletseas, and Wolfgang Thomas, editors, *ICALP (1)*, volume 5555 of *Lecture Notes in Computer Science*, pages 653–664. Springer, 2009.

[37] Daniel Lokshtanov, Daniel Marx, and Saket Saurabh. Known algorithms on graphs of bounded treewidth are probably optimal. pages 777–789, 2011.

[38] László Lovász. On determinants, matchings, and random algorithms. In *FCT*, pages 565–574, 1979.

[39] László Lovász. Communication complexity: a survey. *Paths, flows, and VLSI-Layout*, pages 235–265, 1990.

[40] László Lovász and Michael E. Saks. Communication complexity and combinatorial lattice theory. *J. Comput. Syst. Sci.*, 47(2):322–349, 1993.

[41] Ketan Mulmuley, Umesh V. Vazirani, and Vijay V. Vazirani. Matching is as easy as matrix inversion. 7(1):105–113, 1987.

[42] Rolf Niedermeier. *Invitation to Fixed-Parameter Algorithms*. 2002.

[43] Christos H. Papadimitriou and Mihalis Yannakakis. On limited nondeterminism and the complexity of the V-C dimension. *Journal of Computer and System Sciences*, 53(2):161 – 170, 1996.

[44] Ran Raz and Boris Spieker. On the "log rank"-conjecture in communication complexity. *Combinatorica*, 15(4):567–588, 1995.

[45] Neil Robertson and Paul D. Seymour. Graph minors. III. Planar tree-width. *J. Comb. Theory*, 36(1):49–64, 1984.

[46] Ryan Williams. Finding paths of length k in $\mathcal{O}^{\star}(2^k)$ time. *Inf. Process. Lett.*, 109(6):315–318, 2009.

[47] Virginia Vassilevska Williams. Multiplying matrices faster than Coppersmith-Winograd. In Howard J. Karloff and Toniann Pitassi, editors, *STOC*, pages 887–898. ACM, 2012.

# A  Omitted proofs of Section 2

## A.1  Proof of Proposition 2.2

*Proof.* (sketch) We briefly sketch how to get a nice tree decomposition that is rooted at the introduce edge bag for a chosen edge $\{u, v\} \in E$. To get an empty root bag instead, it suffices to add a series of forget vertex bags that ends in a new (empty) root.

Let $\mathbb{T}$ be a tree decomposition for $G$ and let $\{u, v\} \in E$ be the chosen edge. Pick an arbitrary bag, say $B_r$, of $\mathbb{T}$ that contains both $u$ and $v$ or forget $v$ as the root and follow the arguments given by Kloks [34] to generate a nice tree decomposition with root bag $B_r$ in linear time.

Now additionally do the following. Below each leaf add a series of introduce vertex bags, starting from an empty bag, to make all (new) leaf bags be empty. Regarding the introduce edge bags, first observe that for each vertex there is only one forget vertex bag; this follows from the connectedness property of tree decompositions. To add an introduce edge bag for an edge $\{p, q\}$ take the higher of the forget vertex bags for $p$ and $q$, split the bag into two identical copies and use the lower one to introduce $\{p, q\}$. For the root bag $B_r$, this requires a series of forget vertex bags, until only $u$ and $v$ are left in the highest bag. Then use that bag to introduce $\{u, v\}$. □

# B  Omitted proofs of Section 3

## B.1  Proof of Lemma 3.13

*Proof.* For every $p, q \in \Pi(U)$:

$$
\begin{aligned}
(\mathcal{C}\mathcal{C}^T)[p, q] &= \sum_{(V_1, V_2) \in \mathtt{cuts}(t)} [(V_1, V_2) \sqsubseteq p][(V_1, V_2) \sqsubseteq q] \\
&= \sum_{(V_1, V_2) \in \mathtt{cuts}(t)} [(V_1, V_2) \sqsubseteq p \wedge (V_1, V_2) \sqsubseteq q] \\
&= \sum_{(V_1, V_2) \in \mathtt{cuts}(t)} [(V_1, V_2) \sqsubseteq p \sqcap q] \\
&= 2^{\#\mathtt{blocks}(p \sqcap q) - 1} \\
&\equiv [p \sqcap q = \{U\}].
\end{aligned}
$$

The key here is that we count the number of cuts which are coarser than both $p$ and $q$, i.e., cuts $(V_1, V_2) \in \mathtt{cuts}(t)$ such that each set of $p$ or $q$ is completely contained in $V_1$ or $V_2$. It follows easily, that this is equivalent to each set of $p \sqcap q$ being contained in $V_1$ or $V_2$. Considering the blocks of $p \sqcap q$ it is clear that the block containing 1 must be in $V_1$, by the definition of $\mathtt{cuts}(t)$. Any further block can be either in $V_1$ or $V_2$, which gives a total of $2^{\#\mathtt{blocks}(p \sqcap q) - 1}$ coarser cuts. Clearly, that number is odd if and only if there is a single block in $p \sqcap q$ (the one containing 1), i.e., if $p \sqcap q = \{U\}$. □

## B.2  Proof of Lemma 3.15

*Proof.* First observe that a basis minimizing the weight is just a lexicographicaly minimum basis, as independent sets of rows form a matroid. Notice that we may assume that $n \leq 2m$ since we can start with the first $m$ rows and then iteratively introduce $m$ new rows, find a new basis and discard all introduced rows not in this basis. This takes at most $n/m$ iterations, so if we manage to perform one iteration in $m^\omega$ time, this procedure indeed implements the lemma.

Recall that a matrix is in *row echelon form* if

1. All nonzero rows (rows with at least one 1) are above all rows of all zeroes.

2. The leading coefficient (the first nonzero number from the left, also called the pivot) of a nonzero row is always strictly to the right of the leading coefficient of the row above it.

3. All entries in a column below a leading entry are zeroes (implied by the first two criteria)

We use the folklore fact that an $n \times 2n$ matrix can be brought into row-echelon form that is obtained using only elementary row operations in $\mathcal{O}(n^\omega)$ time (see for example [12, Proposition 16.9]).

We proceed as follows: given an $n \times m$ matrix with $m \leq n \leq 2m$ we first transpose it and then sort the columns on their weight in increasing order. Then we bring it in row-echelon form in $\mathcal{O}(m^\omega)$ as discussed above. Since elementary row operations preserve linear column combinations we can now restrict ourselves to finding a minimum weight basis in the obtained matrix that is in reduced echelon form. And it is easy to see that in this matrix the minimum weight basis is exactly the set of columns that contain a leading 1: it is clearly a set of linearly independent columns, and if a column is not in the basis it can be expressed as a linear combination of columns to the left of it since for all non-zero entries of that column there is a leading 1 appear on the same row in a column before it by the row echelon property. $\square$

### B.3 Proof of Lemma 3.6

*Proof.* In the following let $U$ be a set, let $\mathcal{A}, \mathcal{A}' \subseteq \Pi(U) \times \mathbb{N}$, and assume that $\mathcal{A}'$ represents $\mathcal{A}$. Note that the union and join operations are symmetric and hence we only need to show the preservation of representation with respect to one of the parameters.

**Union:** We have for all $q \in \Pi(U)$ and $\mathcal{B} \subseteq \Pi(U) \times \mathbb{N}$ that

$$\mathtt{opt}(q, \mathcal{A}' \uplus \mathcal{B}) = \min\{\mathtt{opt}(q, \mathcal{A}'), \mathtt{opt}(q, \mathcal{B})\} = \min\{\mathtt{opt}(q, \mathcal{A}), \mathtt{opt}(q, \mathcal{B})\} = \mathtt{opt}(q, \mathcal{A} \uplus \mathcal{B}).$$

**Insert:** We may assume that $X = \{e\}$ since for $Y \cap Z = \emptyset$, $\mathtt{ins}(Y \cup Z, \mathcal{A})$ equals $\mathtt{ins}(Y, \mathtt{ins}(Z, \mathcal{A}))$. Note that if $\{e\} \in q$, $\mathtt{opt}(q, \mathtt{ins}(\{e\}, \mathcal{C})) = \infty$ and representation is preserved trivially. Otherwise, we have for every $\mathcal{C} \subseteq \Pi(U) \times \mathbb{N}$ and any $q \in \Pi(U \cup \{e\})$ with $\{e\} \notin q$ that

$$\mathtt{opt}(q, \mathtt{ins}(\{e\}, \mathcal{C})) = \min\{w \mid (p, w) \in \mathcal{C} \land p_{\uparrow U \cup \{e\}} \sqcap q = \{U \cup \{e\}\}\}$$
$$= \min\{w \mid (p, w) \in \mathcal{C} \land p \sqcap q_{\downarrow U} = \{U\}\}$$
$$= \mathtt{opt}(q_{\downarrow U}, \mathcal{C}).$$

Note that the second equivalence uses the fact that $e$ must be connected to some $e' \in U$ in $q$, since $\{e\} \notin q$. Now preservation of representation follows since

$$\mathtt{opt}(q, \mathtt{ins}(\{e\}, \mathcal{A}')) = \mathtt{opt}(q_{\downarrow U}, \mathcal{A}') = \mathtt{opt}(q_{\downarrow U}, \mathcal{A}) = \mathtt{opt}(q, \mathtt{ins}(\{e\}, \mathcal{A}')).$$

**Shift:** We have for all $q \in \Pi(U)$ and $w' \in \mathbb{N}$ that

$$\mathtt{opt}(q, \mathtt{shft}(w', \mathcal{A}')) = w' + \mathtt{opt}(q, \mathcal{A}') = w' + \mathtt{opt}(q, \mathcal{A}) = \mathtt{opt}(q, \mathtt{shft}(w', \mathcal{A})).$$

**Glue:** We may assume that $a, b \in U$ since otherwise $\mathtt{glue}(ab, \mathcal{A}) = \mathtt{ins}(\{a, b\} \setminus U, \mathtt{glue}(ab, \mathcal{A}))$. Note that for every $\mathcal{C} \subseteq \Pi(U) \times \mathbb{N}$ and any $q \in \Pi(U)$ we have that

$$\mathtt{opt}(q, \mathtt{glue}(ab, \mathcal{C})) = \min\{w \mid (p, w) \in \mathcal{C} \land (p \sqcap U[ab]) \sqcap q = \{U\}\}$$
$$= \min\{w \mid (p, w) \in \mathcal{C} \land p \sqcap (q \sqcap U[ab]) = \{U\}\}$$
$$= \mathtt{opt}(q \sqcap U[ab], \mathcal{C}).$$

Then preservation of representation follows since

$$\mathtt{opt}(q, \mathtt{glue}(ab, \mathcal{A}')) = \mathtt{opt}(q \sqcap U[ab], \mathcal{A}') = \mathtt{opt}(q \sqcap U[ab], \mathcal{A}) = \mathtt{opt}(q, \mathtt{glue}(ab, \mathcal{A})).$$

It is clear from its definition that the variant of Glue with three parameters also preserves representation.

**Project:** We may assume that $X = \{e\}$ since for $Y \cap Z = \emptyset$, $\mathtt{proj}(Y \cup Z, \mathcal{A})$ equals $\mathtt{proj}(Y, \mathtt{proj}(Z, \mathcal{A}))$.
Then for every $\mathcal{C} \subseteq \Pi(U) \times \mathbb{N}$ and any $q \in \Pi(U \setminus \{e\})$ we have that

$$\begin{aligned}
\mathtt{opt}(q, \mathtt{proj}(\{e\}, \mathcal{C})) &= \min \left\{ w \,\middle|\, (p, w) \in \mathcal{C} \wedge p_{\downarrow U \setminus \{e\}} \sqcap q = \{U \setminus \{e\}\} \wedge \{e\} \notin p \right\} \\
&= \min \left\{ w \,\middle|\, (p, w) \in \mathcal{C} \wedge p \sqcap q_{\uparrow U} = \{U\} \right\} \\
&= \mathtt{opt}(q_{\uparrow U}, \mathcal{C}).
\end{aligned}$$

Now preservation of representation follows directly

$$\mathtt{opt}(q, \mathtt{proj}(\{e\}, \mathcal{A}')) = \mathtt{opt}(q_{\uparrow U}, \mathcal{A}') = \mathtt{opt}(q_{\uparrow U}, \mathcal{A}) = \mathtt{opt}(q, \mathtt{proj}(\{e\}, \mathcal{A})).$$

**Join:** We may assume that $\mathcal{A}, \mathcal{B} \subseteq \Pi(\hat{U}) \times \mathbb{N}$ since otherwise we can use

$$\mathtt{join}(\mathcal{A}, \mathcal{B}) = \mathtt{join}(\mathtt{ins}(U' \setminus U, \mathcal{A}), \mathtt{ins}(U \setminus U', \mathcal{B})).$$

For every $\mathcal{B} \subseteq \Pi(\hat{U}) \times \mathbb{N}$ and $r \in \Pi(\hat{U})$ it holds that

$$\begin{aligned}
\mathtt{opt}(r, \mathtt{join}(\mathcal{A}', \mathcal{B})) &= \min \left\{ w_1 + w_2 \,\middle|\, (w_1, p) \in \mathcal{A}', (w_2, q) \in \mathcal{B} \wedge p \sqcap q \sqcap r = \{\hat{U}\} \right\} \\
&= \min \left\{ \mathtt{opt}(q \sqcap r, \mathcal{A}') + w_2 \,\middle|\, (w_2, q) \in \mathcal{B} \right\} \\
&= \min \left\{ \mathtt{opt}(q \sqcap r, \mathcal{A}) + w_2 \,\middle|\, (w_2, q) \in \mathcal{B} \right\} \\
&= \min \left\{ w_1 + w_2 \,\middle|\, (w_1, p) \in \mathcal{A}, (w_2, q) \in \mathcal{B} \wedge p \sqcap q \sqcap r = \{\hat{U}\} \right\} \\
&= \mathtt{opt}(r, \mathtt{join}(\mathcal{A}, \mathcal{B})). \qquad \square
\end{aligned}$$

## B.4 Proof of Theorem 3.18

*Proof.* The algorithm is the following: use the above dynamic programming formulation from Subsection 3.4 to compute $A_r$, but after evaluation of every entry $A_x$, use Theorem 3.7 to store $\mathtt{reducematching}(A_x)$ rather than $A_x$. This is possible because by definition $A_x$ clearly is a set of weighted matchings. Since $\mathtt{reducematchings}(\mathcal{A})$ represents $\mathcal{A}$ and the recurrence uses only the operators defined in Definition 3.2 which all preserve representation by Lemma 3.6, we have as invariant that for every $x \in \mathbb{T}$ the entry stored for $A_x$ represents $A_x$ by Lemma 3.6. In particular, the stored value for $A_r(\mathbf{s})$ dominates $A_r(\mathbf{s})$ and hence we can safely read off the answer to the problem as done in Section 3.4.

Let us focus on the time analysis: since for all operations from Definition 3.2 we can apply Proposition 3.3, the bottleneck clearly is the $\mathtt{reducematchings}$ algorithm. Recall that the guessing of the edge $v_1 v_n$ gives overhead of at most $\mathtt{pw}$ or $\mathtt{tw}$ so this will be subsumed by the last term of the claimed running time.

In the first algorithm that is given a path decomposition, note we can assume $x$ is either a leaf, introduce vertex, forget vertex or introduce edge bag. In this case the size of $A_x(\mathbf{s})$ is by our invariant at most $2^{|s^{-1}(1)|/2}$. Then the time needed to perform $\texttt{reducematchings}(A_x(\mathbf{s}))$ is at most $2^{\omega/2|s^{-1}(1)|}\texttt{pw}^{\mathcal{O}(1)}$ and the time to compute $A_x(\mathbf{s})$ for every $\{0,1,2\}^{B_x}$ is

$$\sum_{\mathbf{s}\in\{0,1,2\}^{B_x}} 2^{\omega/2|s^{-1}(1)|}\texttt{pw}^{\mathcal{O}(1)} \leq \sum_{i_0+i_1+i_2=|B_x|} \binom{|B_x|}{i_0,i_1,i_2} 1^{i_0} 1^{i_2} (2^{\omega/2})^{i_1} \texttt{pw}^{\mathcal{O}(1)}$$
$$= (2+2^{\omega/2})^{|B_x|}\texttt{pw}^{\mathcal{O}(1)},$$

where the last equality is due to the multinomial theorem. Hence $A_x$ can be computed for every $x \in \mathbb{T}$ in the claimed time bound.

For tree decompositions, note that if $x$ is a join bag then

$$|A_x(\mathbf{s})| \leq \sum_{\mathbf{l}+\mathbf{r}=\mathbf{s}} 2^{|l^{-1}(1)|/2} 2^{|r^{-1}(1)|/2} = \prod_{i=1}^{|B_x|} \sum_{l+r=s_i} 2^{[l=1]/2+[r=1]/2} = (2^{3/2})^{|s^{-1}(1)|} 4^{|s^{-1}(2)|}.$$

where we use in the second-last equation that

$$\sum_{l_i+r_i=s_i} 2^{[l_i=1]/2+[r_i=1]/2} = \begin{cases} 1 & \text{if } s_i = 0 \\ 2^{3/2} & \text{if } s_i = 1 \\ 4 & \text{if } s_i = 2. \end{cases} \tag{3}$$

Using this, it can also be seen that the reduce operation on $A_x(\mathbf{s})$ is performed in time bounded by $(2^{3/2})^{|s^{-1}(1)|} 4^{|s^{-1}(2)|} 2^{(\omega-1)/2|s^{-1}(1)|} |B_x|^{\mathcal{O}(1)}$. Then the total time needed to evaluate $A_x(\mathbf{s})$ for every $\mathbf{s} \in \{0,1,2\}^{B_x}$ is bounded by

$$\sum_{i_0+i_1+i_2=|B_x|} \binom{|B_x|}{i_0,i_1,i_2} 1^{i_0} (2^{(\omega+2)/2})^{i_1} 4^{i_2} = (5+2^{(\omega+2)/2})^{|B_x|},$$

and the claimed running time follows by the multinomial theorem. □